\documentclass[runningheads]{llncs}
\usepackage[T1]{fontenc}
\usepackage{tikz}
\usepackage{graphicx}
\usepackage{amssymb}
\usepackage{amsmath}
\usepackage{program}
\usepackage{url}
\usepackage{caption}
\usepackage{subcaption}
\usepackage{wrapfig}
\usepackage{listings}
\usepackage{scalerel}
\usepackage{array}
\usepackage{diagbox}
\usepackage{booktabs}
\usepackage{upgreek}
\usepackage{hyperref}
\usepackage{cleveref}
\usepackage{pgfplots}  
\pgfplotsset{width=6.6cm,compat=1.7}  
\usepackage{color}
\definecolor{RYB2}{RGB}{30,144,255}
\definecolor{RYB1}{RGB}{178,34,34}
\definecolor{RYB3}{RGB}{128,0,128}
\definecolor{RYB4}{RGB}{34,139,34}

\DeclareMathOperator*{\argmin}{arg\,min}

\usepackage{multirow}
\usepackage{verbatim}

\begin{document}

\title{Adnexal Mass Segmentation with Ultrasound Data Synthesis}
% Matt: Improving Adnexal Mass Segmentation in Imbalanced Ultrasound Data with Image Synthesis (?)
% Accurate Ovarian Lesion Segmentation under Imbalanced Ultrasound Data
% Clara's: Adnexal Mass Segmentation with Imbalanced Ultrasound Data
%
\titlerunning{Adnexal Mass Segmentation with Ultrasound Data Synthesis}
% % If the paper title is too long for the running head, you can set
% % an abbreviated paper title here
\author{Clara Lebbos\inst{1} \and Jen Barcroft\inst{1} \and Jeremy Tan\inst{1} \and Johanna P. M\"uller\inst{2} \and Matthew Baugh\inst{1} \and  Athanasios Vlontzos\inst{1} \and Srdjan Saso\inst{1}   \and Bernhard Kainz\inst{1,2} }
% %
\authorrunning{C. Lebbos et al.}
\institute{\inst{1}Imperial College London
\email{clara.lebbos18 $\vert$ srdjan.saso01 $\vert$ b.kainz@imperial.ac.uk} \\
\inst{2}Friedrich-Alexander-Universität Erlangen-N\"urnberg
% \url{***}}
}

\institute{Imperial College London, London, UK \\ \email{clara.lebbos18 $\vert$ srdjan.saso01 $\vert$ b.kainz@imperial.ac.uk}  \and
Friedrich-Alexander-Universität Erlangen-N\"urnberg, Erlangen, Germany
}
\maketitle 

\begin{abstract}
Ovarian cancer is the most lethal gynaecological malignancy. The disease is most commonly asymptomatic at its early stages and its diagnosis relies on expert evaluation of transvaginal ultrasound images. Ultrasound is the first-line imaging modality for characterising adnexal masses, it requires significant expertise and its analysis is subjective and labour-intensive, therefore open to error. Hence, automating processes to facilitate and standardise the evaluation of scans is desired in clinical practice. Using supervised learning, we have demonstrated that segmentation of adnexal masses is possible, however, prevalence and label imbalance restricts the performance on under-represented classes. To mitigate this we apply a novel pathology-specific data synthesiser. We create synthetic medical images with their corresponding ground truth segmentations by using Poisson image editing to integrate less common masses into other samples. Our approach achieves the best performance across all classes, including an improvement of up to 8\% when compared with nnU-Net baseline approaches.
\end{abstract}

\section{Introduction}

One in two people will develop cancer, with ovarian cancer being the most lethal gynaecological malignancy \cite{nhs_cancer_stat}. Around 75\% of women diagnosed with ovarian cancer are already at the advanced stages of the disease due to the earlier stages often showing no symptoms \cite{ovarian_cancer_stats_2}, leading to a survival rate of only $35\%$ \cite{cancerresearch}. %Standard treatment for diagnosed women includes a total abdominal hysterectomy (TAH) and bilateral salpingo-oophorectomy (BSO) to remove the womb, ovaries and fallopian tubes, subsequently making the women infertile if premenopausal \cite{cancerresearch}. In addition, up to $18\%$ of women experience ovarian cysts which could be benign or malignant \cite{stats_ovarian_cyst}, and a false diagnosis puts women at risk of becoming infertile. 

Current diagnostic methods for detecting ovarian cancer include tumour markers (CA-125) and the analysis of a transvaginal ultrasound image. Ultrasound classification of adnexal masses can be a challenging process. Current ultrasound based diagnostic models exist, but still rely on the extraction of ultrasound features, which is time-consuming, labor-intensive, operator-dependant and subject to error. Accurate classification of adnexal masses is necessary to reduce incorrect diagnosis and unnecessary surgeries. With an efficient workflow, population screening for ovarian cancer could be feasible, leading to earlier disease detection and increased survival rates. Evidently, there is tangible value in developing a tool able to automatically extract the features in transvaginal ultrasounds accurately and consistently.

Supervised Deep Learning is a fast-growing area of research that has been successfully applied in the medical field to predict illnesses, design treatments, and distinguish between benign and malignant masses \cite{hesamian2019deep}. For ovarian cancer diagnosis, Deep Learning could be used to automatically segment transvaginal ultrasound features and extract characteristics required by risk models such as the number of cysts and their dimensions.

A major challenge of ovarian cancer diagnosis is the accurate interpretation of ultrasound images. Although ultrasound is the imaging modality of choice, as it is accessible, non-invasive, and does not use radiation \cite{746626}, it is also ``notorious for having significant noise with a low signal-to-noise ratio'' \cite{roscoecoping}. This represents a challenge for both medical experts and segmentation algorithms. The wide variety in shape and size of adnexal masses can also pose challenges for segmentation.

% \subsection*{Objectives and Challenges}

In this paper we automate and regularise ultrasound feature extraction, with the aim of improving the accuracy of diagnosis and reducing unnecessary referrals.
We focus on transvaginal ultrasound images and segment four adnexal mass features: (1) lesions, (2) locules, (3) solid areas and (4) papillations -- this has not yet been attempted in literature to the best of our knowledge. The presence or absence, number of occurrences, and dimensions of adnexal masses represent the ultrasound features used by risk models to determine the probability of malignancy.
Additionally, some adnexal masses are more common than others, and some of the less common masses are the most difficult to detect, even for medical experts. This can lead to an imbalanced dataset and a poor segmentation of under-represented masses. 

\noindent\textbf{Contributions:}
Current Deep Learning methods struggle to accurately segment images when trained on imbalanced datasets.
Our contributions are as follows:
1) we propose a novel method to synthesise segmentation data with Poisson image editing~\cite{perez2003poisson} to mitigate class imbalance;
2) we extensively evaluate our method on transvaginal ultrasound data and compare it to domain experts, which we show to be highly biased in an inter-observer study;
3) models trained with our additional generated data outperform the current state-of-the-art nnU-Net~\cite{nn_unet} (+3.7\% DSC and +8\% DSC for the augmented classes), achieving excellent and high-standard segmentations for lesions and locules $\geq 0.94$ DSC, and good quality segmentation for solid areas and papillations $\approx 0.82$ DSC.

\noindent\textbf{Related Work:} 
The task of segmenting transvaginal ultrasound scans has previously been attempted by researchers in the medical field. There are models that segment polycystic ovary ultrasounds \cite{kumar2014segmentation}, or detect various features to give a direct risk probability prediction and bypass risk models \cite{jin2021multiple,budd2019confident}. Direct risk prediction is a black box giving clinicians less insight, control and confidence over the diagnosis in contrast to a segmentation-based approach. To our knowledge, no model has been developed to extract the specific ultrasound features~\cite{timmerman2008simple} required by robust, externally validated ultrasound-based risk models such as Assessment of Different Neoplasias of Adnexa (ADNEX) and Simple Rules, developed by the International Ovarian Tumour Analysis group ~\cite{no2011management,RMI_comparison,timmerman2008simple,timmerman2005logistic,van2014evaluating,chudecka2015roma}. The goal of these guidelines and clinical algorithms is to determine whether the masses are benign or malignant, with a percentage probability of risk. Their parameters are mostly based on time-consuming and operator-dependant assessment of the presence, appearance, and size of pathological structures in ultrasound images.

Synthesising images, specifically medical images, has been a growing area of research. Deep Learning models have been developed to synthesise images from one medical modality to another \cite{jiao2020self}, or entirely new ultrasound images \cite{cronin2020using,tan2022detecting,chotzoglou2021learning,meng2019weakly,meng2020mutual}. The latter does not guarantee that pathological textures are preserved or that synthesised images are medically plausible and realistic. This could lead to generated ultrasound images with implausible ovarian features overlapping. GAN-based methods have also been explored for data anonymisation, augmentation~\cite{shin2018medical}, and domain transfer~\cite{gholami2018novel}. These methods are usually not tailored towards a specific disease.
\section{Methodology}

\begin{figure}[ht]
    \centering
    \includegraphics[width=\textwidth, height=4.5cm]{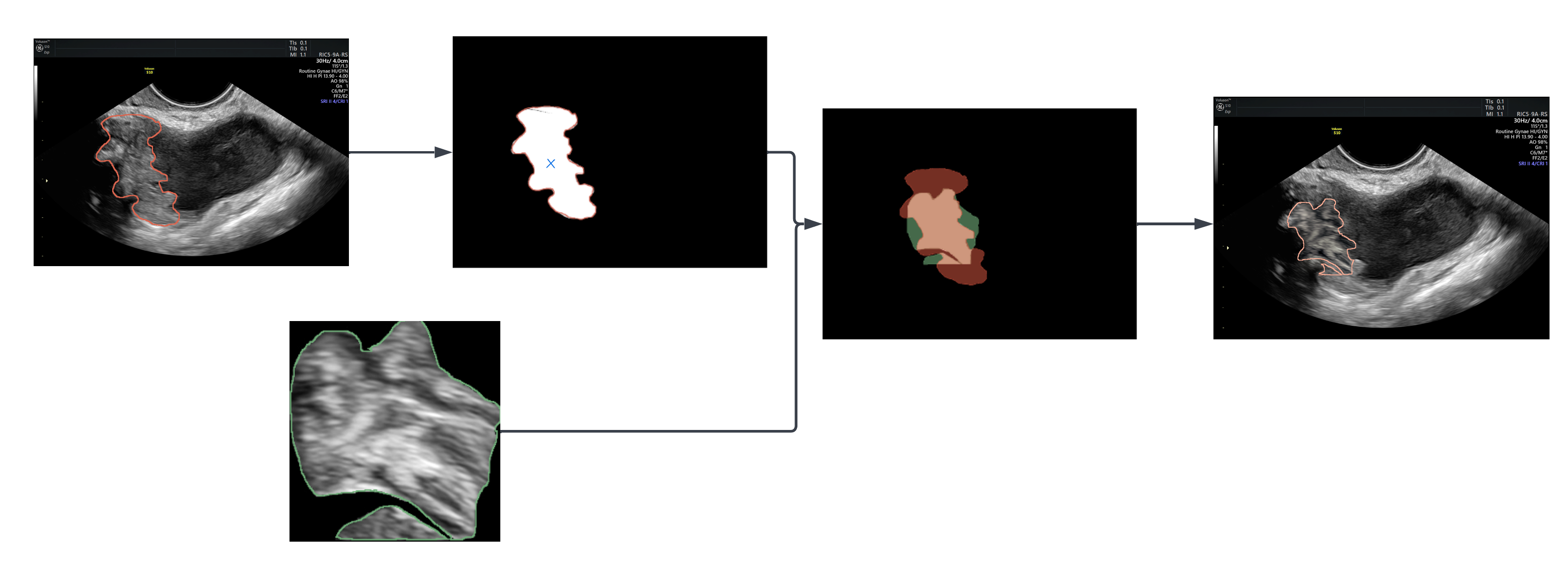}
    \caption{Overview of the Synthesiser Implementation: extracted papillations (green mask) from existing images are randomly positioned and seamlessly blended into target images with a solid area (red mask) by performing the intersection of the masks to ensure they are overlapping}
    \label{im:synthesiser}
\end{figure}

Deep neural networks such as nnU-Net \cite{nn_unet} require a large amount of data to perform well. This requirement is commonly addressed with extensive data augmentation techniques such as rotating and scaling. However, this does not improve cases of imbalanced datasets. %Our novel data synthesiser can assist with this, generating new, high quality images of underrepresented classes and their corresponding ground truth annotations for supervised models to train on.
%Our method relies upon high quality blending of under-represented classes into new images.
% The synthesiser can be used to increase the size of the dataset for classes which do not have enough data by extracting them from existing images and blending them into others.
In this work we aim to synthesise new data by extracting examples of less-seen masses and integrating them into other samples in which such features would be clinically plausible. 
To do this we cannot na\"ively copy and paste the masses from one image to another, as this would result in unrealistic image discontinuities. This simple technique was first trialled and resulted in a significant decrease of performance for solid areas and papillations (-2.7\% and -3.3\% DSC respectively). Instead we propose a novel method of synthesising data based on the Poisson image editing algorithm \cite{perez2003poisson}.
This method allows us to balance the dataset with good quality imaging, and subsequently improve the performance of the model, specifically for the augmented classes. An overview of our approach is shown in Figure~\ref{im:synthesiser}.

\noindent\textbf{Data Synthesis of Solid Areas and Papillations:} 
%Our goal is to increase the occurrence of underrepresented classes -papillations and solid areas- in our dataset in a way that is realistic and anatomically plausible.
%In order to do so 
We first use ground truth annotations to extract and sort papillations with respect to their size and relative position to solid areas.
Following that, we identify target images that do not contain any of our upsampled classes nor any class that would render the co-occurrence with the upsampled class medically implausible.
A papillation is then randomly selected and blended into the target image. 

Each chosen papillation is aligned with the solid area of the target image and then offset by a third of the solid area's width. The direction of the offset is chosen such that it matches the relative position of the papillation in the original image. It is a physiological requirement for papillations to overlap with solid areas.
Poisson image editing \cite{perez2003poisson,opencvpoisson} is then used to seamlessly blend the papillation patch onto the target image. 
%The masks of the solid areas, papillations and their intersections are subsequently generated, seen in Figure \ref{im:intersection}.
The final step is to modify the ground truth segmentation by adding the classification of pixels for the source image in the region of interest, such that the segmentation contains the extracted mass.

Since this method performs an intersection of the papillation and solid area, the synthesised images are not complete duplicates of the existing data, while keeping the texture of the masses. This means that the solid areas are also modified which we hypothesise makes the model more robust to this pathology as well as for papillations.

% \begin{figure}[htp]
%   \begin{center}
%     \includegraphics[width=3cm, height=2cm]{implementation/images/label4extract.png}
%   \end{center}
%   \caption{Extracted papillation from image}
%     \label{img:label4}
% \end{figure}

% \begin{figure}[h]
% \centering
% \resizebox{0.8\textwidth}{!}{
%   \begin{subfigure}[b]{0.28\textwidth}
%     \includegraphics[width=\textwidth]{implementation/images/label3mask.png}
%     \caption{Mask of the solid area class \newline}
%   \end{subfigure}
%   \hfill
%   \begin{subfigure}[b]{0.28\textwidth}
%     \includegraphics[width=\textwidth]{implementation/images/label4mask.png}
%     \caption{Mask of the papillation class \newline}
%     \label{mask2}
%   \end{subfigure}
%   \hfill
%   \begin{subfigure}[b]{0.28\textwidth}
%     \includegraphics[width=\textwidth]{implementation/images/label1mask.png}
%     \caption{Intersection of the solid area and the papillation masks (ROI)}
%     \label{im:intersection}
%   \end{subfigure}}
%   \caption{Intersection of the solid area and the papillation masks for sample masses}
% \end{figure}

\noindent\textbf{Poisson Image Editing:} 
Poisson image editing~\cite{perez2003poisson} blends the source image into the target image by making edits in the gradient domain.
For a source image $g$ and the destination image $f^*$, we want to calculate an interpolant $f$ over a region $\Omega$ representing the pixels under the mask with a boundary $\delta \Omega$ such that the squared error between the gradients of $f$ and $g$ is minimised, whilst having equal intensity values to the destination image at the region boundaries (Eqn.~\ref{eqn:poissonproblem}).
The solution is then the Poisson equation Eqn.~\ref{eqn:poissonsolution}.
\begin{equation}
    f = \argmin_f \iint_{\Omega} \|\nabla f - v\|^2 ; f\Big\vert_{\partial\Omega} = f^*\Big\vert_{\partial\Omega} ,  v = \nabla g
    \label{eqn:poissonproblem}
\end{equation}

\begin{equation}
   \Updelta f = \text{div } v \text{ over } \Omega \text{, with } f\Big\vert_{\partial\Omega} = f^* \Big\vert_{\partial\Omega}
   \label{eqn:poissonsolution}
\end{equation}

\begin{figure}[htb]
\centering

  \begin{subfigure}[b]{.48\textwidth}
    \begin{subfigure}[b]{0.48\textwidth}
      \includegraphics[width=1.0\textwidth]{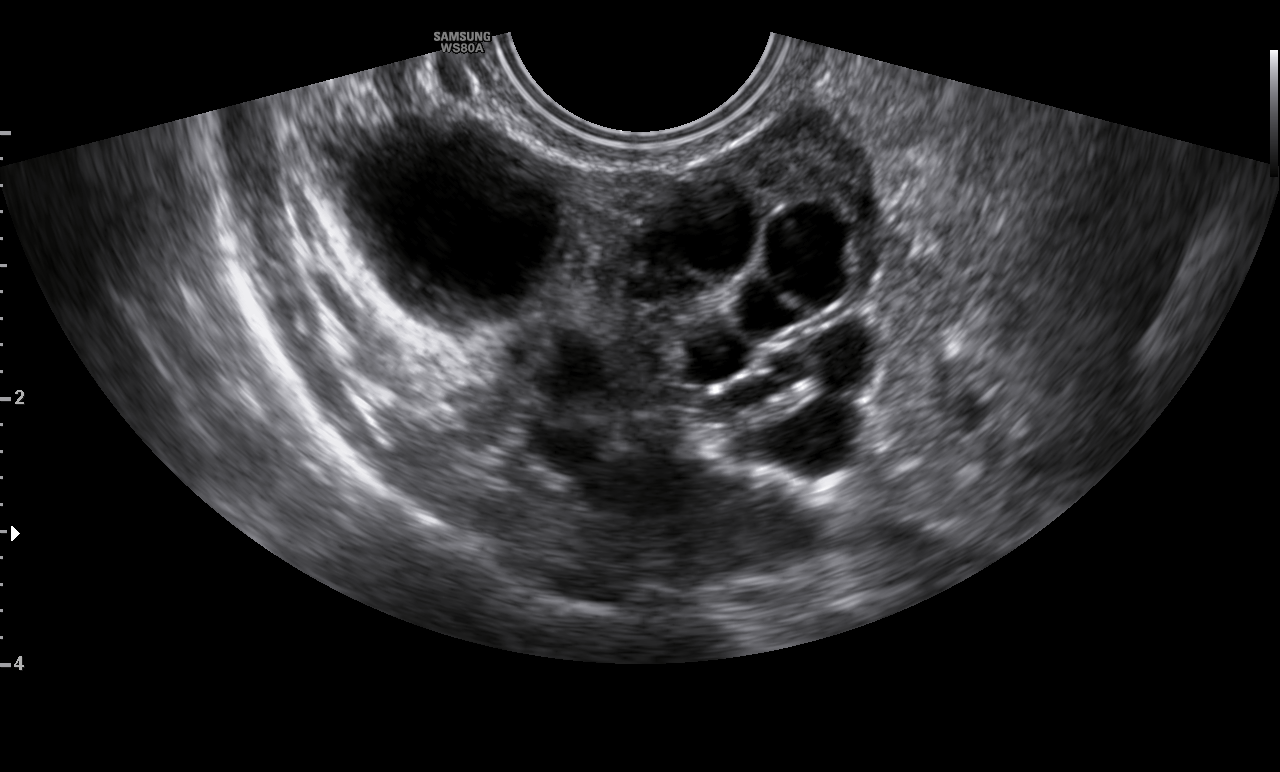}
    \end{subfigure}
    %\hfill
    \begin{subfigure}[b]{0.48\textwidth}
        \includegraphics[width=1.0\textwidth]{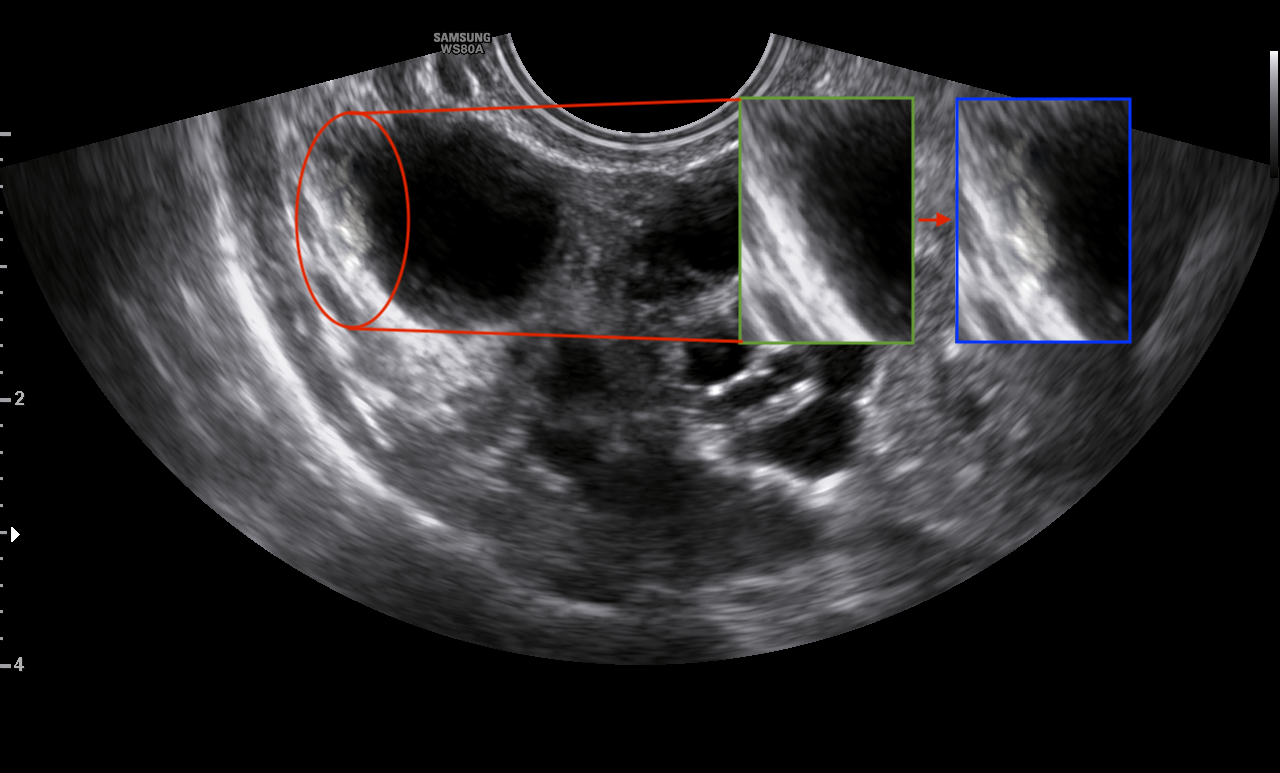}
    \end{subfigure}
    \caption{Realistic Samsung W10 example}
  \end{subfigure}
  \hfill
  \begin{subfigure}[b]{.48\textwidth}
    \begin{subfigure}[b]{0.48\textwidth}
      \includegraphics[width=\textwidth]{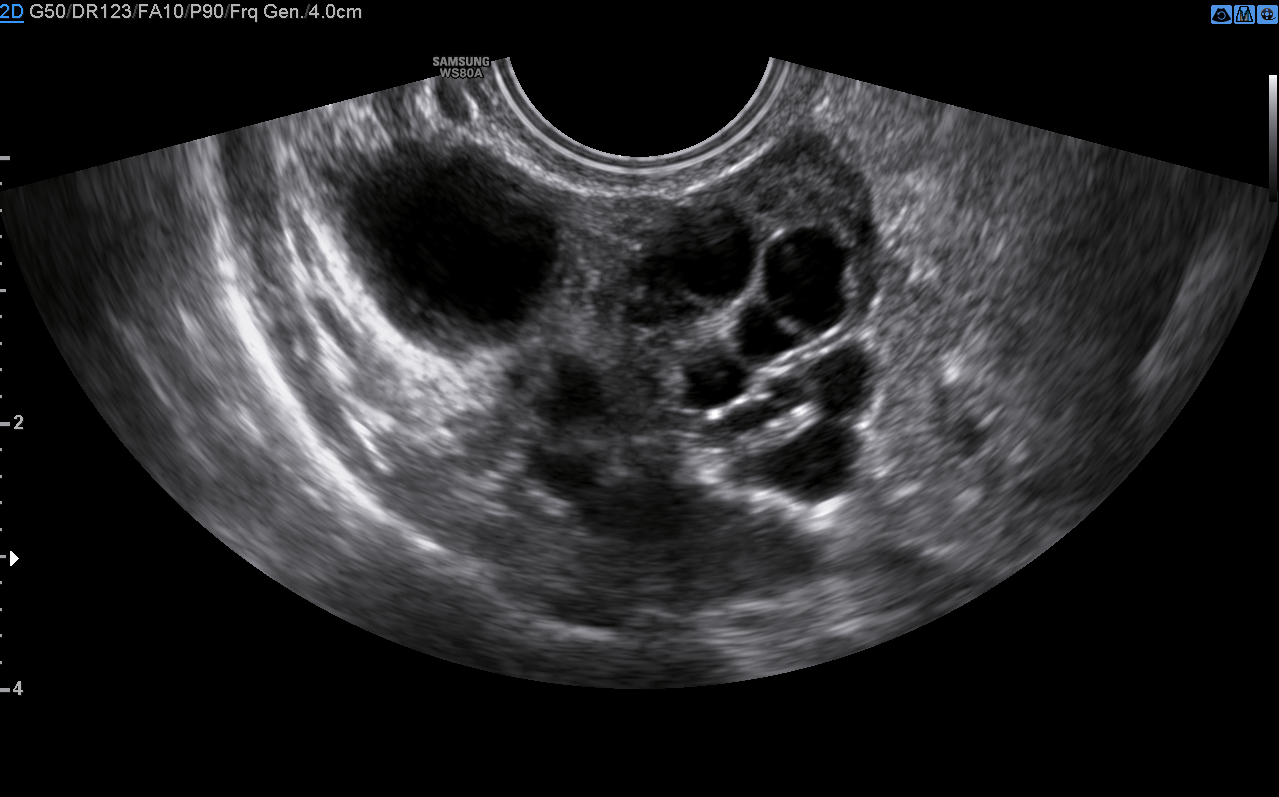}
    \end{subfigure}
%   \hfill
    \begin{subfigure}[b]{0.48\textwidth}
      \includegraphics[width=\textwidth]{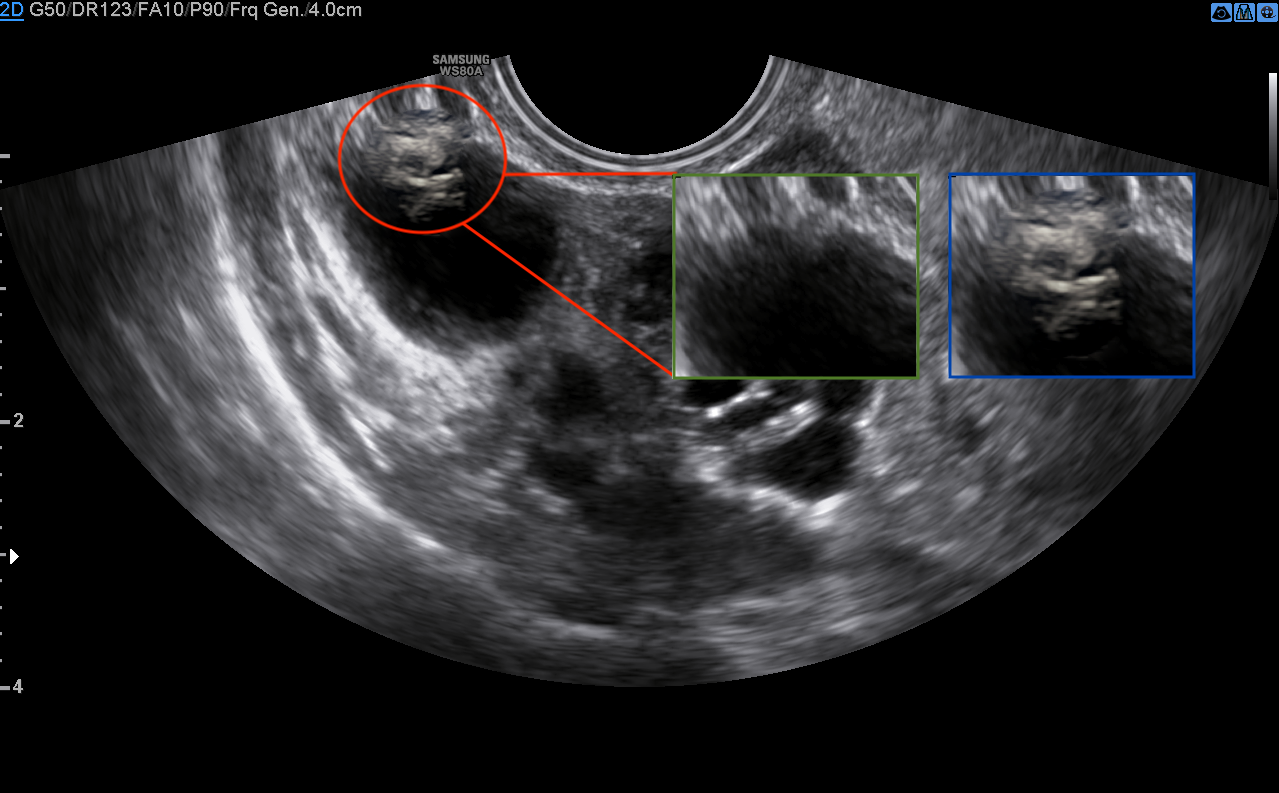}
    \end{subfigure}
    \caption{Less realistic Samsung W10 example}
  \end{subfigure}

  \begin{subfigure}[b]{.48\textwidth}
    \begin{subfigure}[b]{0.48\textwidth}
      \includegraphics[width=1.0\textwidth]{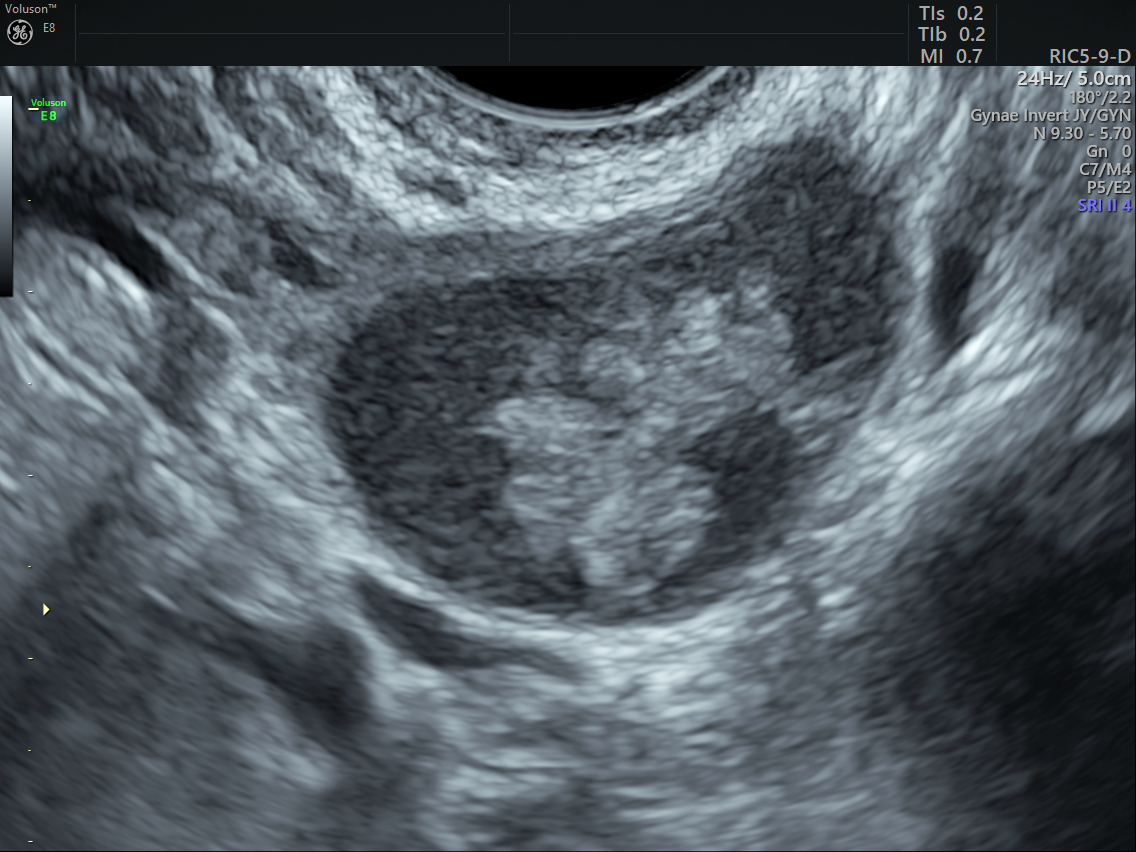}
    \end{subfigure}
   %\hfill
    \begin{subfigure}[b]{0.48\textwidth}
      \includegraphics[width=1.0\textwidth]{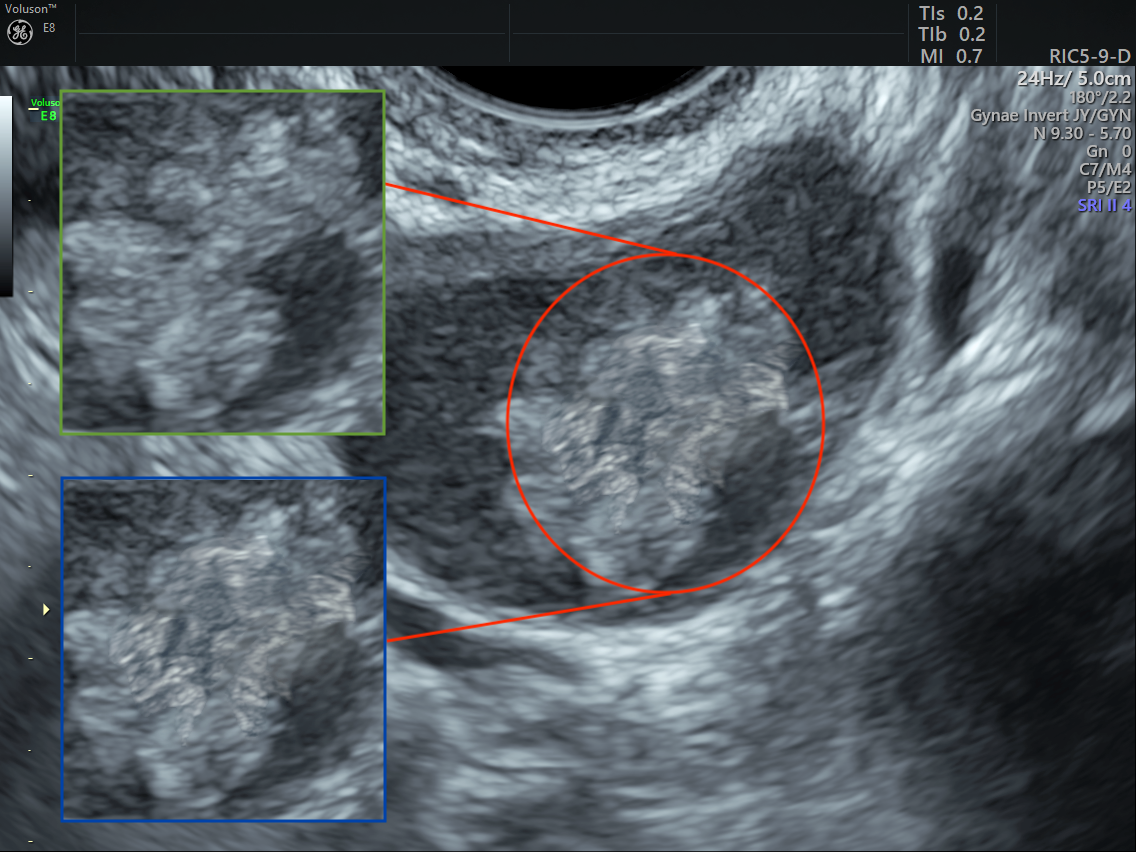}
    \end{subfigure}
    \caption{Realistic Voluson E8 example}
  \end{subfigure}
  \hfill
  \begin{subfigure}[b]{.48\textwidth}
    \begin{subfigure}[b]{0.48\textwidth}
      \includegraphics[width=\textwidth]{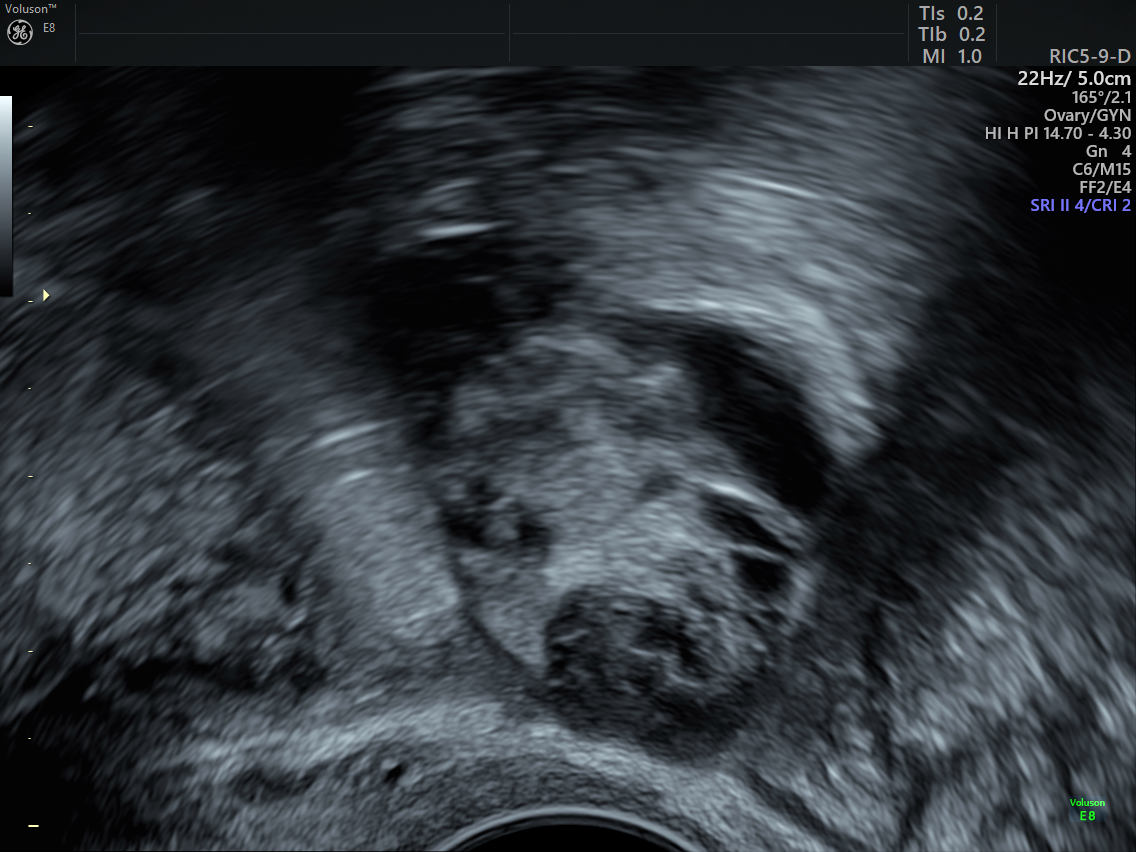}
    \end{subfigure}
    % \hfill
    \begin{subfigure}[b]{0.48\textwidth}
      \includegraphics[width=\textwidth]{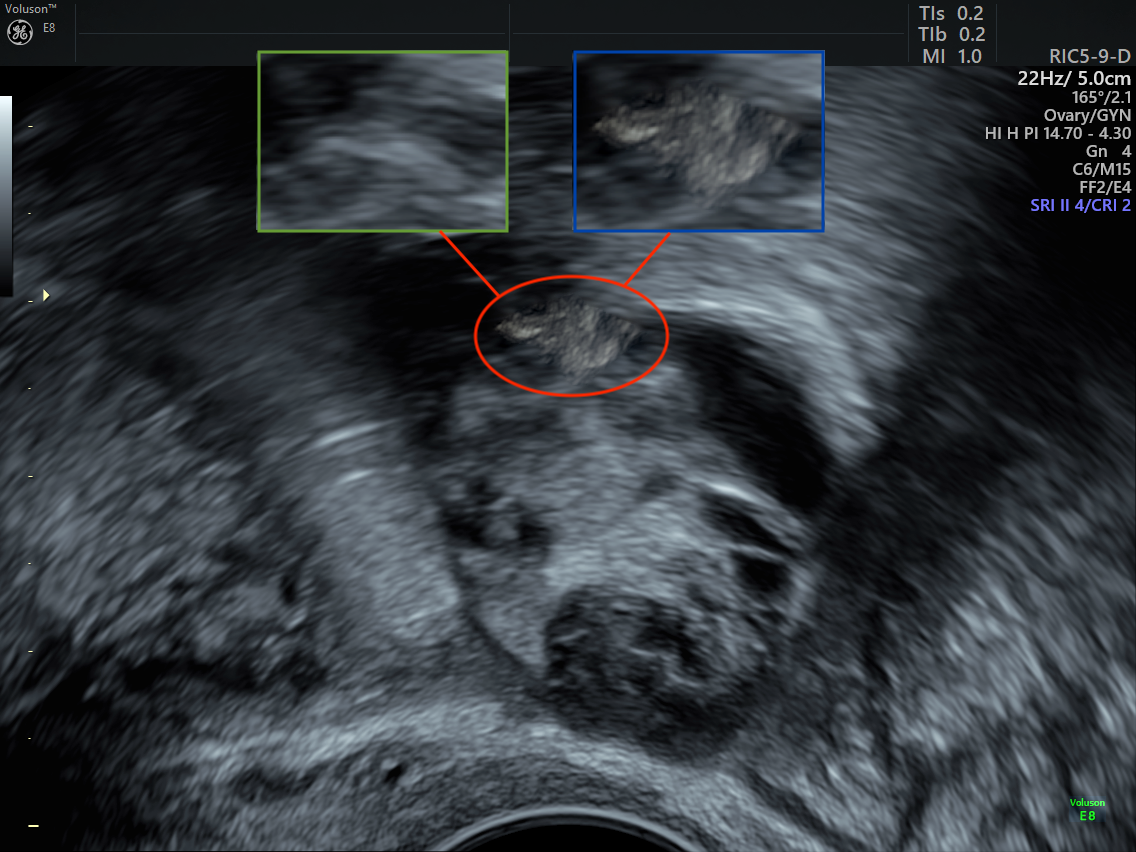}
    \end{subfigure}
    \caption{Less realistic Voluson E8 example}
  \end{subfigure}
  
  \caption{Original and synthesised ultrasound image pairs. On the synthesised image of each the green boxes is the original image area and the blue box is the modified region of interest.}
  \label{implementation:synthesised_image_examples}
\end{figure}

Our synthesiser takes $v$ as the gradient of the source image, rather than using mixed gradients \cite{perez2003poisson}, so that the texture of the extracted region is maintained. Selected example outputs for this approach are shown for the (expert-rated) best and worst artificially generated signs of pathology in Fig.~\ref{implementation:synthesised_image_examples}.

% \begin{figure}[h]
% \centering
% \resizebox{0.8\textwidth}{!}{
%   \begin{subfigure}[b]{0.5\textwidth}
%     \includegraphics[width=\textwidth]{implementation/images/leftmost.png}
%     \caption{Left position}
%   \end{subfigure}
%   \hfill
%   \begin{subfigure}[b]{0.5\textwidth}
%     \includegraphics[width=\textwidth]{implementation/images/top.png}
%     \caption{Top position}
%   \end{subfigure}}
%   \caption{Sample of mask of a solid area (in yellow) with the position for the centre of the papillation (in purple)}
%   \label{im:positions}
% \end{figure}

% \begin{figure}[h]
% \centering
% \resizebox{\textwidth}{!}{
%   \begin{subfigure}[b]{0.5\textwidth}
%     \includegraphics[width=\textwidth]{implementation/images/generate4_2og.png}
%     \caption{Original ultrasound image\newline \newline}
%   \end{subfigure}
%   \hfill
%   \begin{subfigure}[b]{0.5\textwidth}
%     \includegraphics[width=\textwidth]{implementation/images/generate4_2.png}
%     \caption{Synthesised image with added papillation. Green box is the original image and Blue box is the modified region of interest}
%   \end{subfigure}}
%   \caption{Synthesised ultrasound images with their corresponding original image classified as less realistic by experts}
%   \label{im:synth_image}
% \end{figure}

\section{Evaluation and Results}

\begin{figure}
\centering
\begin{minipage}{.46\textwidth}
    \centering
    \resizebox{0.95\textwidth}{!}{
      \begin{subfigure}[c]{0.5\textwidth}
        \includegraphics[width=\textwidth]{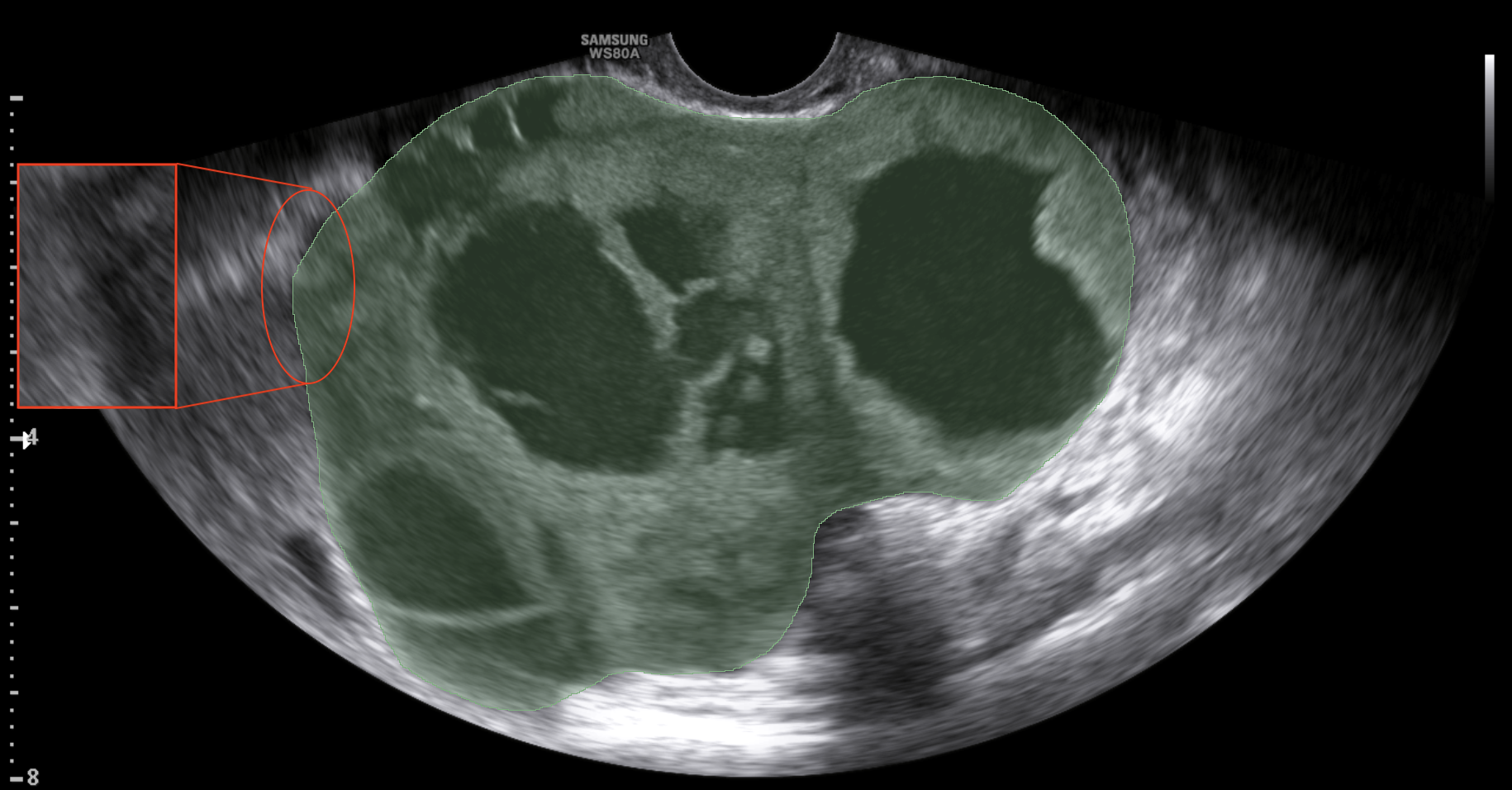}
        \caption{Lesion segmentation}
        \label{im:label1zoom}
      \end{subfigure}
      \hfill
      \begin{subfigure}[c]{0.5\textwidth}
        \includegraphics[width=\textwidth]{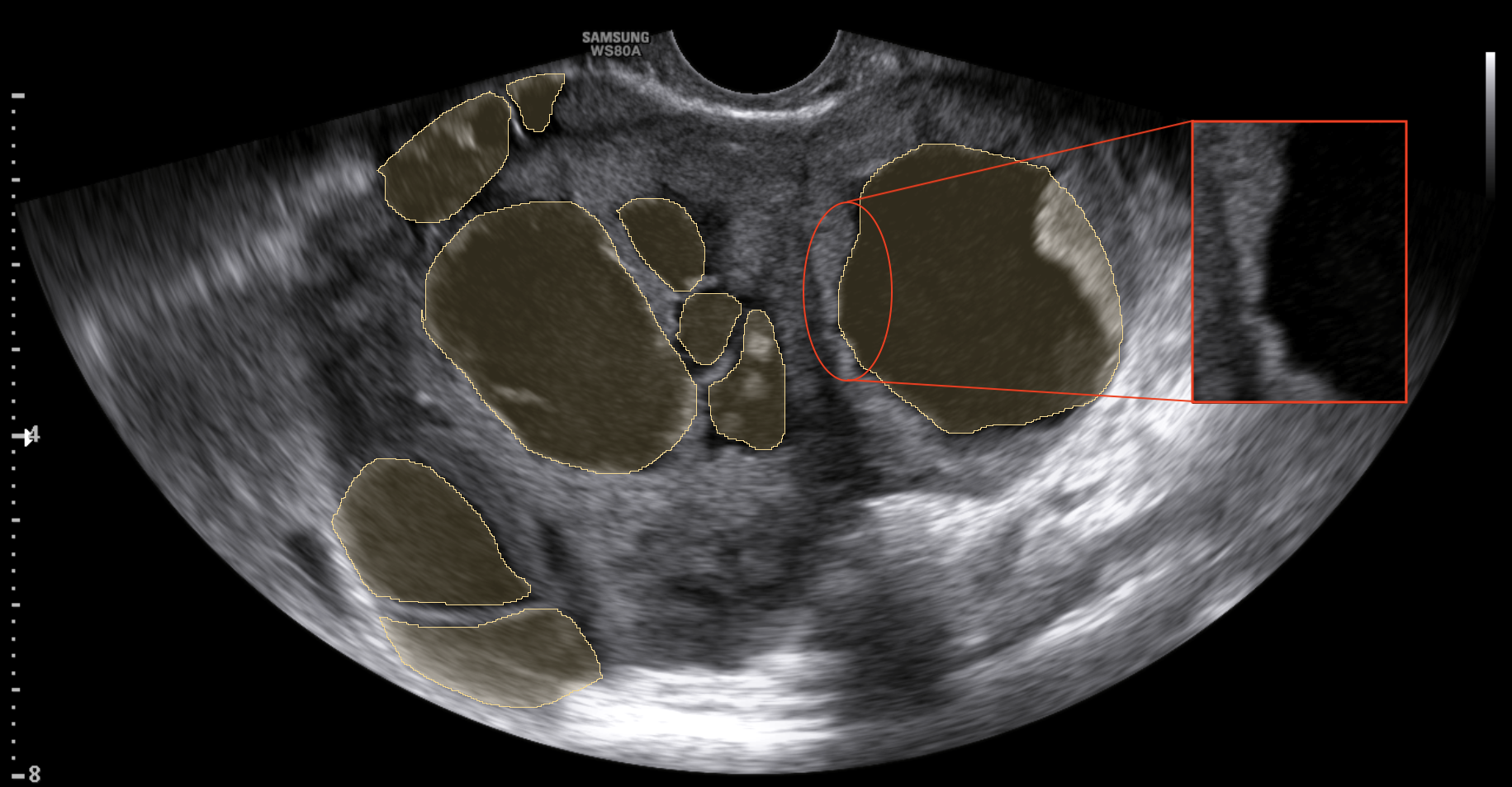}
        \caption{Locule segmentation}
        \label{im:label2zoom}
      \end{subfigure}}
    % % \end{figure}
    
    % \begin{figure}[htp]
    \centering
    \resizebox{0.95\textwidth}{!}{
      \begin{subfigure}[b]{0.5\textwidth}
        \includegraphics[width=\textwidth]{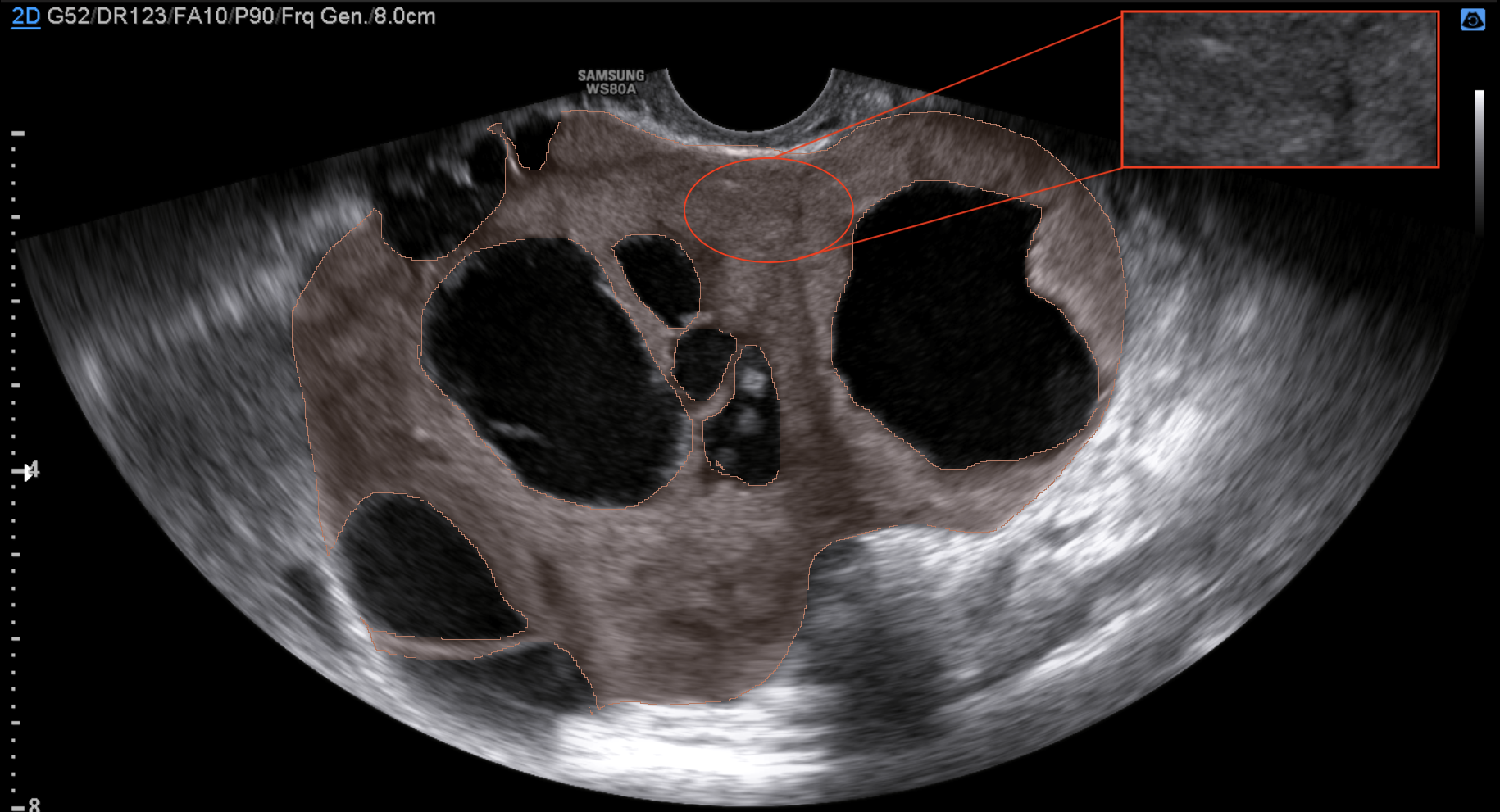}
        \caption{Solid area segmentation}
        \label{im:label3zoom}
      \end{subfigure}
      \hfill
      \begin{subfigure}[b]{0.5\textwidth}
        \includegraphics[width=\textwidth]{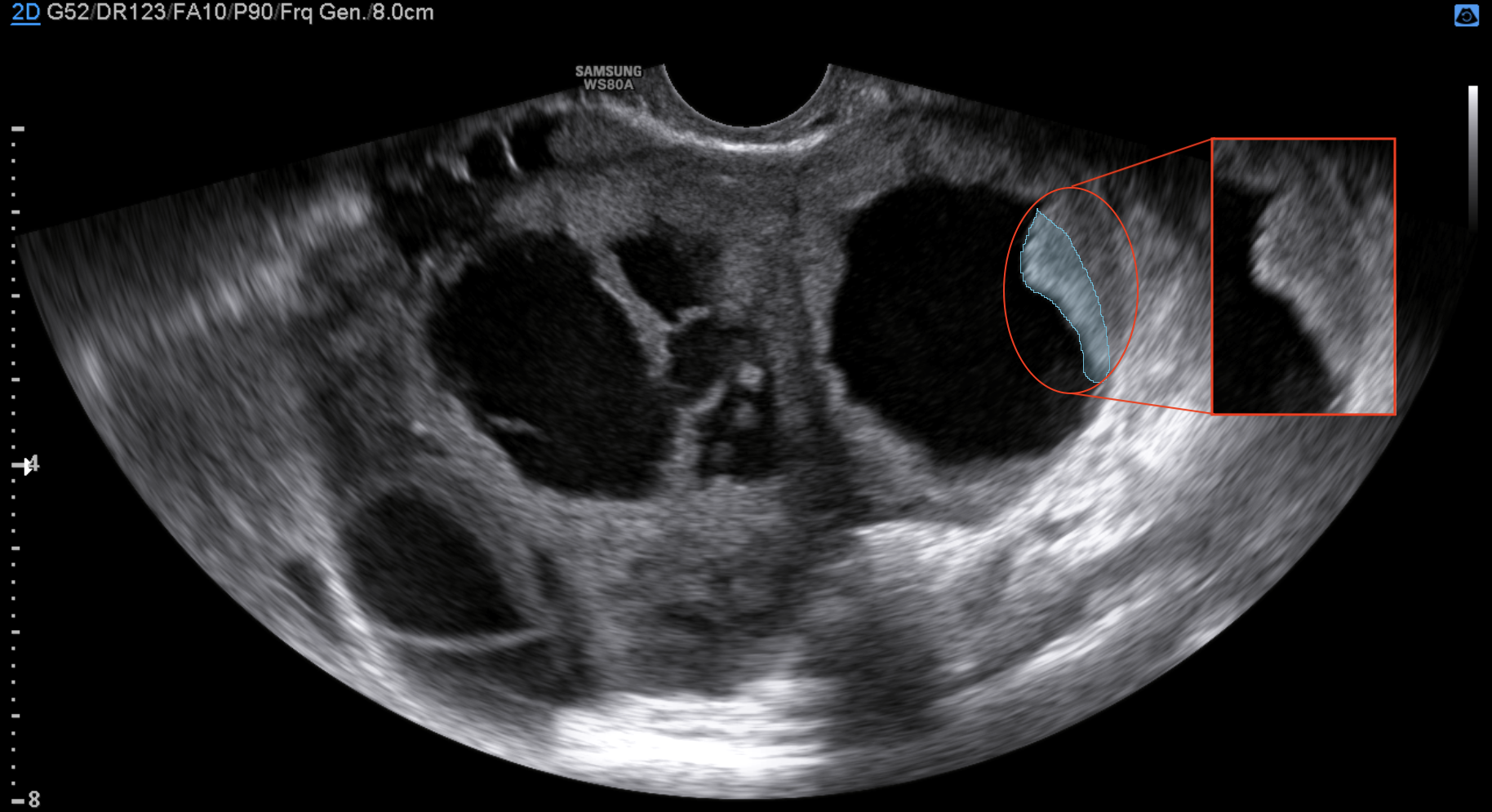}
        \caption{Papillation segmentation}
        \label{im:label4zoom}
      \end{subfigure}}
      
      \caption{Ultrasound image showing examples of each adnexal mass}
      \label{fig:ultrasound_examples}
\end{minipage}%
\qquad
\begin{minipage}{.46\textwidth}
    \centering
    \begin{tikzpicture}
    \begin{axis}[xbar,
    height = 0.95\textwidth,
    width = 0.93\textwidth,
    % y axis line style = { opacity = 0 },
    axis lines = left,
    % tickwidth         = 0pt,
    enlarge y limits  = 0.2,
    xmin = 0,
    xmax = 750,
    % enlarge x limits  = 0.02,
    symbolic y coords = {Papillation, Solid area, Locule, Lesion},
    legend pos = south east,
    nodes near coords,
  ]
  \addplot coordinates { (166,Papillation)         (225,Solid area)
                         (498,Locule)  (532,Lesion) };
  \addplot coordinates { (282,Papillation)         (345,Solid area)
                         (616,Locule)  (650,Lesion) };
    \legend{$D_1$, $D_2$}

    \end{axis}
    \end{tikzpicture}
    \caption{Comparison of the class balance: original dataset $D_1$ and extended dataset with synthesised images $D_2$.}
    \label{fig:dataset_balance_compare}
\end{minipage}
\end{figure}

\noindent\textbf{Dataset:} We have curated a transvaginal ultrasound dataset for this study using a GE Voluson E10, E8, S10 Expert or a Samsung Hera W10, WS80A scanner with a transvaginal convex transducer. The dataset contains 532 2D images from 222 patients with their corresponding ground truth segmentations of lesions, locules, solid areas and papillations  (Fig.~\ref{fig:ultrasound_examples}). Segmentation masks have been generated by an expert and verified/corrected by another clinical expert. 
Solid areas and papillation labels are severely under-represented and also naturally less common and hard to detect due to their small size.
We have conducted an inter-observer variability study where three experienced doctors individually segmented adnexal masses in 20 ultrasound images. Consistently for all pairs of observers, the segmentation of solid areas and papillations appeared to be the most challenging for medical professionals (up to 12.7\% decrease). We use the DSC mean and variance across observers as the human-level performance benchmark, and the tolerance for the Surface DSC is based on the variance of this study.

%The dataset is made up of 222 patients with at least one ovarian cyst. The corresponding images of ultrasound scans are .TIF files, do not have a fixed size, and are not isotropic. As demonstrated in Figure \ref{fig:data_distribution_d2}, all images contain at least one lesion. 
% However, other masses are not a requirement for the study and create an imbalance. This dataset imbalance will have to be addressed during training as solid areas and papillations are usually small and harder to detect as they can be mistaken for noise, and the dataset only has 166 images containing a papillation.

% \begin{figure}[htp]
%     \centering
%     \resizebox{0.6\textwidth}{0.22\textheight}{%
%     \begin{tikzpicture}[x={(.01,0)}]
%     \foreach  \l/\x/\c[count=\y] in {Papillation/166/RYB1, 
%     Solid area/225/RYB2, 
%     Locule/498/RYB3, 
%     Lesion/532/RYB4}
%     {\node[left] at (0,\y) {\l};
%     \fill[\c] (0,\y-.4) rectangle (\x,\y+.4);
%     \node[right] at (\x, \y) {\x};}
%     \draw (0,0) -- (600,0);
%     \foreach \x in {100, 200, ..., 600}
%     {\draw (\x,.2) -- (\x,0) node[below] {\x};}
%     \draw (0,0) -- (0,4.5);
%     \end{tikzpicture}}
%     \caption{Distribution of dataset $D_2$ per label}
%   \label{fig:data_distribution_d2}
%   \caption{Data distribution of datasets $D_1$ and $D_2$ per label}
% \end{figure}

We experiment with training models on a dataset of only real images $D_1$ and the same dataset extended with images produced by our data synthesiser $D_2$. The training took approximately 40 hours per fold.
We evaluated each model with the same held-out test set consisting of 53 real samples, ensuring consistency and fairness when comparing the methods.
None of these test images were included in any part of the synthesis process when creating $D_2$.
As a baseline we used nnU-Net \cite{nn_unet} with region-based training to allow for overlapping labels.

We have synthesised ultrasound images with lesions, locules, solid areas and papillations. In our case, a patient can only have a papillation if there is also a solid area, a locule and a lesion present, which represents only 10\% of our dataset. This means that the dataset cannot be entirely balanced, but we have increased the number of images with solid areas and papillations from 225 to 345 and 166 to 282 respectively (Fig.~\ref{fig:dataset_balance_compare}).

\noindent\textbf{Metrics:} Segmentation performance is evaluated between ground truth and model prediction by Dice similarity coefficient (DSC), Surface Dice similarity coefficient (SDSC), Hausdorff Distance at 95th percentile (HD95) and recall.
For images in the test set where one label was not present in both the ground truth and the predicted segmentations, the DSC and SDSC and recall are set to 1 if no false positive prediction is present, \emph{i.e.}, an empty label mask is predicted. To determine whether the improvement of methods from the baseline is significant, we perform a statistical significance test for the DSC and SDSC. We will carry out a two-sided paired t-test on two models trained on the same test set. From this statistical significance test, a p-value is calculated and is compared to a threshold $\alpha$.

% \begin{figure}[h]
%     \centering
%   \begin{subfigure}[b]{0.5\textwidth}
%     \resizebox{1\textwidth}{0.25\textheight}{%
%     \begin{tikzpicture}[x={(.01,0)}]
%     \foreach  \l/\x/\c[count=\y] in {Original Papillation/166/RYB1, Papillation/282/RYB1, 
%     Original Solid area/225/RYB2,
%     Solid area/345/RYB2, 
%     Original Locule/498/RYB3,
%     Locule/616/RYB3,
%     Original Lesion/532/RYB4,
%     Lesion/650/RYB4}
%     {\node[left] at (0,\y) {\l};
%     \fill[\c] (0,\y-.4) rectangle (\x,\y+.4);
%     \node[right] at (\x, \y) {\x};}
%     \draw (0,0) -- (700,0);
%     \foreach \x in {100, 200, ..., 700}
%     {\draw (\x,.2) -- (\x,0) node[below] {\x};}
%     \draw (0,0) -- (0,8.5);
%     \end{tikzpicture}}
%   \end{subfigure}
%   \caption{Distribution of dataset $D_2$ per label}
%     \label{fig:data_distribution_d1_new}
% \end{figure}

% Our results report the mean and standard derivation across the five folds of the Dice similarity coefficient (DSC), Jaccard Index (JI), Hausdorff Distance at the 95th percentile (HD95), and the recall. The standard derivation represents the mean of the standard derivation of each cross validation fold. This allows us to evaluate the performance of the models on the entire training sets. For the qualitative analysis, we use the ensemble of the five models.

%\subsection{Results Details}

\begin{table}[htb]
\centering
\resizebox{\textwidth}{!}{
\begin{tabular}{@{}p{0.5cm}lllll@{}}
\toprule
& \textbf{Model}   & \textbf{DSC}           & \textbf{SDSC}          & \textbf{Recall}        & \textbf{HD95}          \\ \midrule

%\multicolumn{5}{l}{\textbf{Lesions}} \\ \midrule

\multirow{5}{*}{\rotatebox{90}{\textbf{Lesions}}} &  \textbf{BL}             & 0.951 $\pm$ 0.063          & 0.961 $\pm$ 0.005          & 0.948 $\pm$ 0.080          & 10.35 $\pm$ 12.71          \\
& \textbf{BL + BN + DA}     & 0.955 $\pm$ 0.058          & 0.966 $\pm$ 0.006          & 0.947 $\pm$ 0.085          & 9.461 $\pm$ 12.79          \\
& \textbf{BL + S}           & 0.955 $\pm$ 0.056          & 0.965 $\pm$ 0.011          & \textbf{0.955} $\pm$ \textbf{0.067} & 9.379 $\pm$ 12.73           \\
& \textbf{BL + S + BN + DA} & $\textbf{0.957} \pm \textbf{0.050}^{*}$ & \textbf{0.966} $\pm$ \textbf{0.001}          & 0.953 $\pm$ 0.075          & \textbf{9.068} $\pm$ \textbf{11.32} \\
& \textbf{Expert} & $0.98 \pm 0.062$          & -          & $0.975 \pm 0.092$          & $17.88 \pm 55.267$          \\
\midrule

%\multicolumn{5}{l}{\textbf{Locule}} \\ \midrule
\multirow{5}{*}{\rotatebox{90}{\textbf{Locule}}} &
\textbf{BL}               & 0.936 $\pm$ 0.073          & 0.984 $\pm$ 0.062          & 0.934 $\pm$ 0.095          & 10.057 $\pm$ 11.65         \\
& \textbf{BL + BN + DA}     & 0.938 $\pm$ 0.074         & 0.988 $\pm$ 0.047          & 0.93 $\pm$ 0.105           & 9.549 $\pm$ 11.52          \\
& \textbf{BL + S}    & $\textbf{0.943} \pm \textbf{0.064}^*$ & \textbf{0.989} $\pm$  \textbf{0.05} & \textbf{0.944 $\pm$ 0.08} & \textbf{8.852} $\pm$ \textbf{10.17} \\
& \textbf{BL + S + BN + DA} & 0.941 $\pm$ 0.071          & 0.988 $\pm$ 0.045          & 0.936 $\pm$ 0.100          & 9.298 $\pm$ 10.79          \\ 
& \textbf{Expert} & $0.939 \pm 0.229$          & -          & $0.943 \pm 0.229$          & $4.204 \pm 17.1$          \\\midrule

%\multicolumn{5}{l}{\textbf{Solid area}} \\ \midrule
\multirow{5}{*}{\rotatebox{90}{\textbf{Solid area}}} &
\textbf{BL}               & 0.736 $\pm$ 0.383          & 0.852 $\pm$ 0.339         & 0.822 $\pm$ 0.320          & 26.011 $\pm$ 18.93          \\
& \textbf{BL + BN + DA}     & $0.812 \pm 0.330^{\bullet}$          & $0.901 \pm 0.276^{*}$         & 0.854 $\pm$ 0.291          & 26.748 $\pm$ 20.98          \\
& \textbf{BL + S}           & $\textbf{0.816} \pm \textbf{0.327}^{\circ}$ & $\textbf{0.908} \pm \textbf{0.276}^{\circ}$ & \textbf{0.875 $\pm$ 0.265} & \textbf{23.957} $\pm$ \textbf{19.77}          \\
&  \textbf{BL + S + BN + DA} & $0.809 \pm 0.334^{\circ}$          & $0.898 \pm 0.292^{*}$         & 0.864 $\pm$ 0.283          & 25.432 $\pm$ 21.70          \\ 
& \textbf{Expert} & $0.853 \pm 0.345$          & -          & $0.908 \pm 0.27$         & $69.611 \pm 164.532$          \\\midrule

%\multicolumn{5}{l}{\textbf{Papillation}} \\ \midrule
\multirow{5}{*}{\rotatebox{90}{\textbf{Papillation}}} &
\text{BL}               & 0.761 $\pm$ 0.395          & 0.854 $\pm$ 0.325         & 0.805 $\pm$ 0.363         & \textbf{31.746 $\pm$ 26.67}          \\
& \textbf{BL + BN + DA}     & $0.796 \pm 0.376^*$          & 0.843 $\pm$ 0.353        & \textbf{0.815 $\pm$ 0.363} & 36.436 $\pm$ 30.77 \\

& \textbf{BL + S}          & $0.798 \pm 0.380^*$ & $0.878 \pm 0.30^{\circ}$ & 0.814 $\pm$ 0.357         & 37.467 $\pm$ 27.1          \\
& \textbf{BL + S + BN + DA} & $\textbf{0.809} \pm \textbf{0.391}^*$          & $\textbf{0.882} \pm \textbf{0.357}^{\circ}$          & 0.807 $\pm$ 0.371         & 40.851 $\pm$ 28.48          \\
& \textbf{Expert} & $0.873 \pm 0.33$          & -          & $0.954 \pm 0.202$         & $25.318 \pm 66.985$          \\
\bottomrule
\\
\end{tabular}}
\caption{Evaluation and ablation study for DSC, SDSC, Recall and HD95 (in px) scores across models for locules, lesions, solid areas and papillations [\textbf{$\text{BL}$}: Baseline nnU-Net \cite{nn_unet} with region-based training, \textbf{BN}: Batch normalisation, \textbf{DA}: Intense data augmentation, \textbf{S}: With synthesised data] on dataset $D_1$ and $D_2$ for synthesiser. Significance compared to the baseline ($\text{BL}$):  $p^*$ < 0.05, $p^{\circ}$ < 0.01, $p^{\bullet}$ < 0.001)}
\label{tab:results}
\end{table}

%\noindent\textbf{Discussion} 

\noindent\textbf{Results:} Figure \ref{implementation:synthesised_image_examples} shows synthesised images with their corresponding original ultrasound image. For most images, the added papillation is blended seamlessly in the image; the edges are not apparent and the intensities are coherent whilst still keeping the texture of papillations. The less realistic images arise from incompatible textures in the source and destination images. As observed, the textures are obviously different for the less realistic images leading to apparent edges. We sought an evaluation by experienced ultrasound examiners in Gynaecology who rated a selection of images. 

In Table~\ref{tab:results} we observe that the nnU-Net \cite{nn_unet} baseline model, which uses region-based training, performs very well for lesions and locules (>0.9 DSC) but worse for solid areas and papillations (<0.8 DSC). We have also trained the baseline model with more intensive data augmentation and batch normalisation, which improves the segmentation performance. %For all masses, performance improve, demonstrating a better performance of the model for all labels. 
% The recall is lower for lesions and locules but still remains within a standard deviation of the baseline which isn't a significant difference (Tables \ref{tab:part2label1}, \ref{tab:part2label2}). This could be due to the already high recall values for lesion and locule segmentation. On the other hand, the recall value for solid areas and papillations for BL were relatively low (0.822 and 0.805 respectively) and increased by 3.2\% and 1\% respectively (Tables \ref{tab:part2label3}, \ref{tab:part2label4}). The SDSC and DSC also improved considerably for solid areas and papillations, with an increase of 5\% for the SDSC of solid areas and 3.5\% for the DSC of papillations. A paired t-test shows that this model performs significantly better for solid areas and papillations with a p-value under 0.05. Finally, the Hausdorff distance appears to increase for solid areas and papillations, but we believe this is because the model is detecting and segmenting more areas.

The models trained with synthesised data appear to be the best performing for all classes. 
The most significant effect is observed in solid areas where all of the metrics have improved, with the DSC and SDSC increasing by 8\% and 5.6\% respectively, making it significantly better than the baseline with p < 0.01 and considerably better than all other models.
Finally, the performance of the segmentation of papillations is also improved with the DSC and SDSC increasing by 2.1\% and 2.4\% respectively, making the model significantly better than the baseline with regards to the SDSC (p < 0.01). However, the variance and value of the HD95 has increased. We believe this is due to the fragmented nature of this label and more papillations being segmented compared to the other models. The recall is higher which confirms our hypothesis that targeted lesion synthesis can improve ultrasound image segmentation. 

\begin{figure}[htp]
\resizebox{\textwidth}{!}{
\begin{tabular}{ >{\centering\arraybackslash}m{5cm}  >{\centering\arraybackslash}m{5cm}  >{\centering\arraybackslash}m{5cm}  >{\centering\arraybackslash}m{5cm} }

\multicolumn{4}{c}{\textbf{\Large Solid Areas}} \\[.4cm] 

\includegraphics[width=0.33\textwidth, height=40mm]{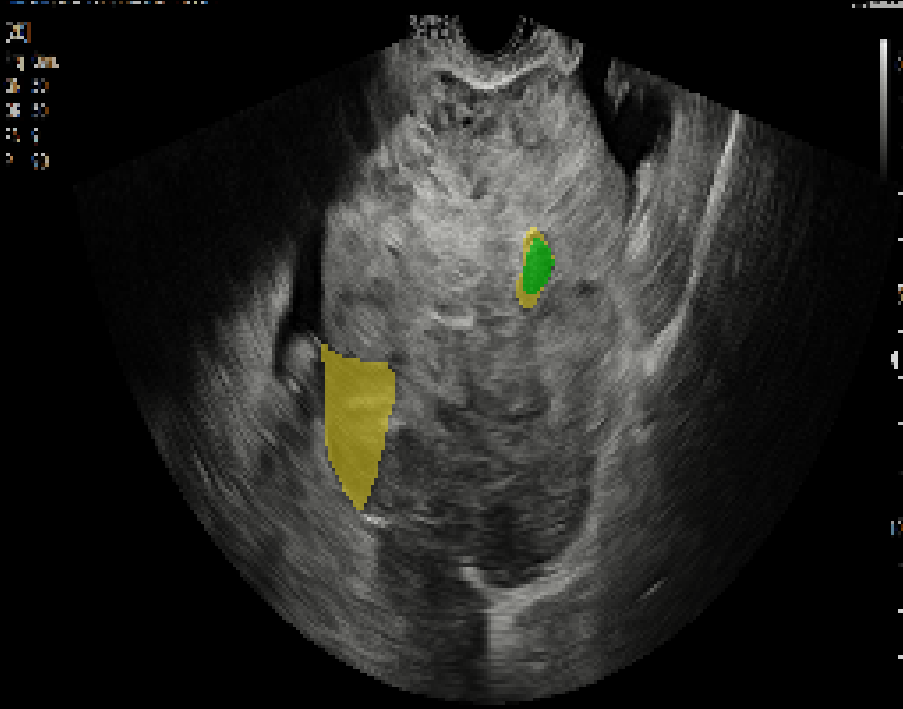} &
\includegraphics[width=0.33\textwidth, height=40mm]{evaluation/images/better26428_0011L1-9.png} &
\includegraphics[width=0.33\textwidth, height=40mm]{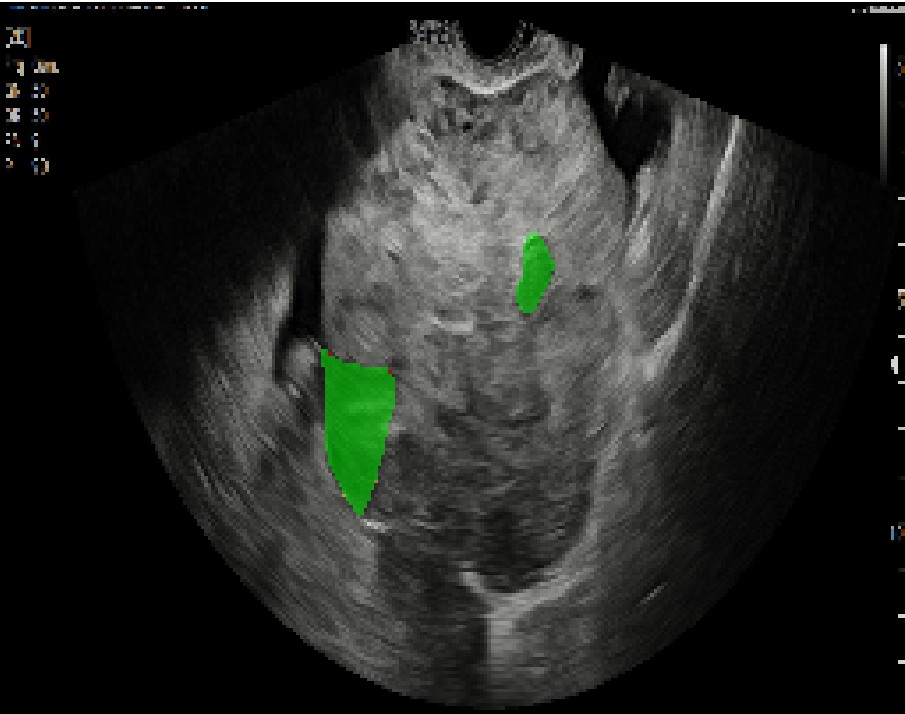} & 
\includegraphics[width=0.33\textwidth, height=40mm]{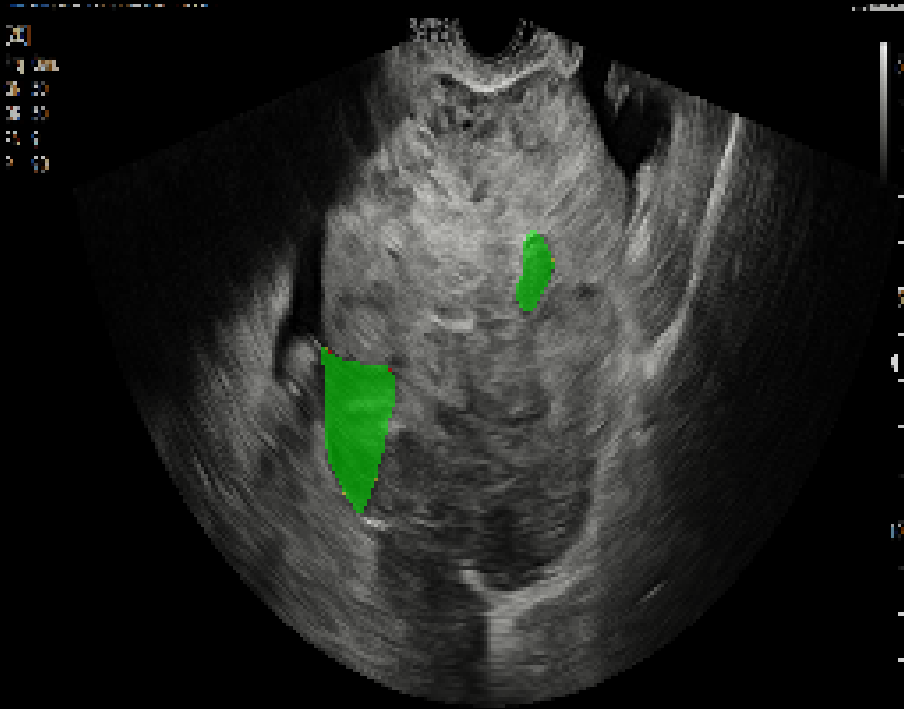} \\

\includegraphics[width=0.33\textwidth, height=40mm]{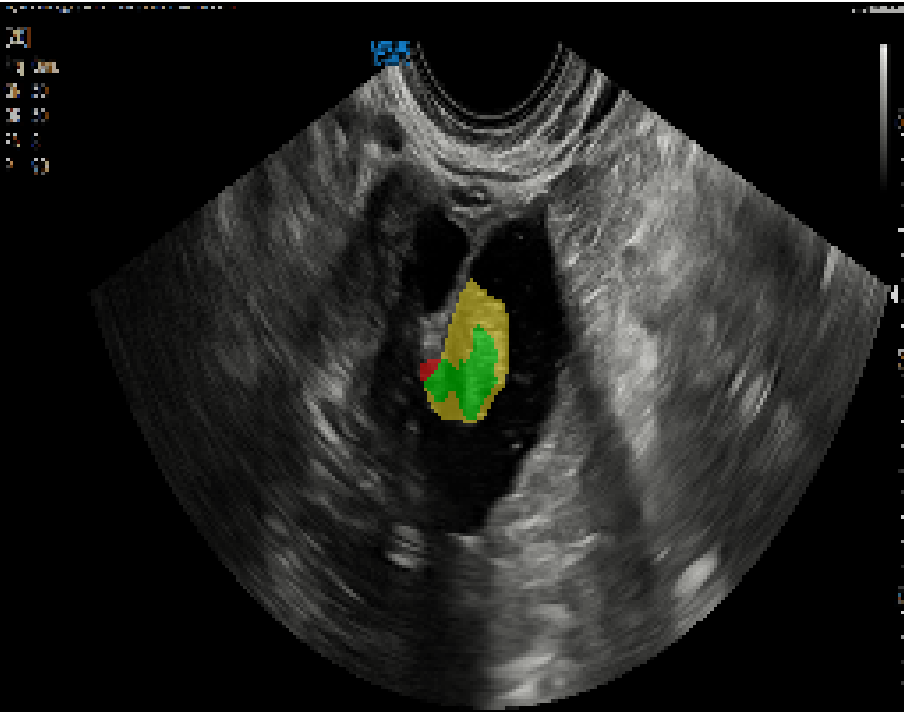} &
\includegraphics[width=0.33\textwidth, height=40mm]{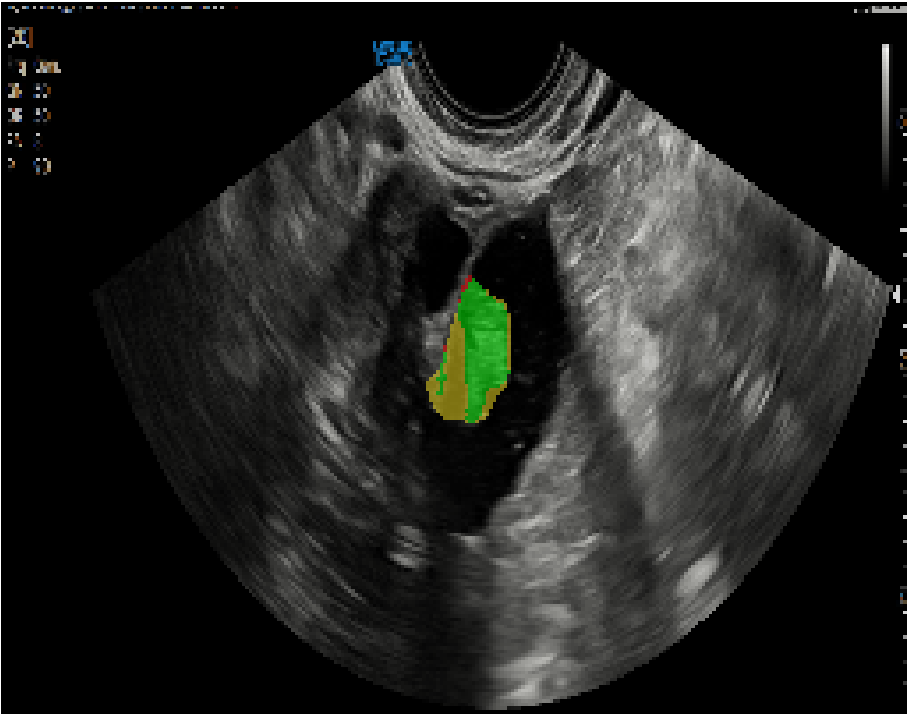} &
\includegraphics[width=0.33\textwidth, height=40mm]{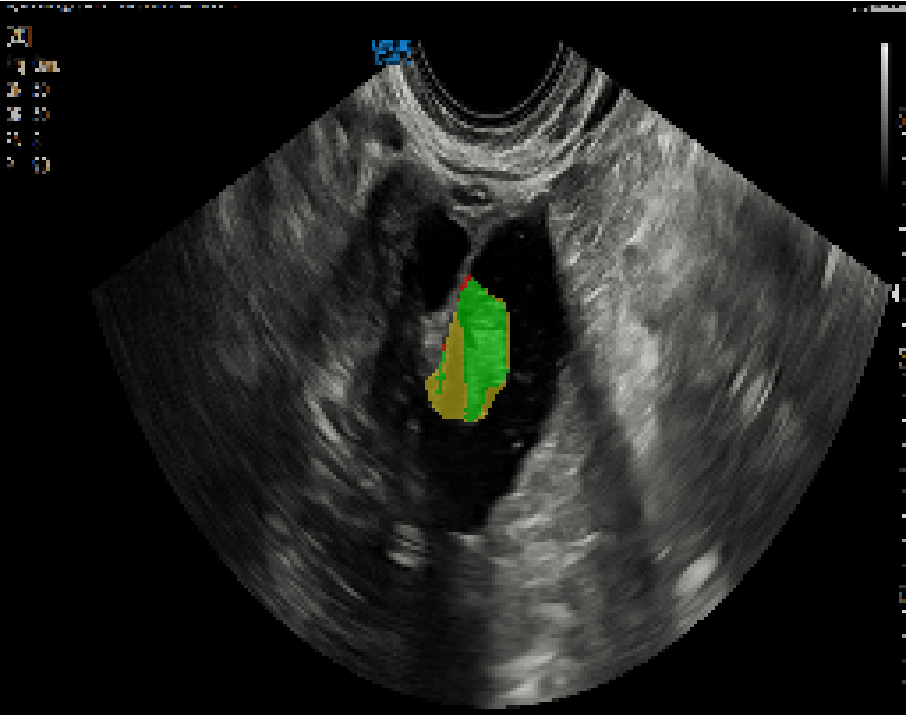} & 
\includegraphics[width=0.33\textwidth, height=40mm]{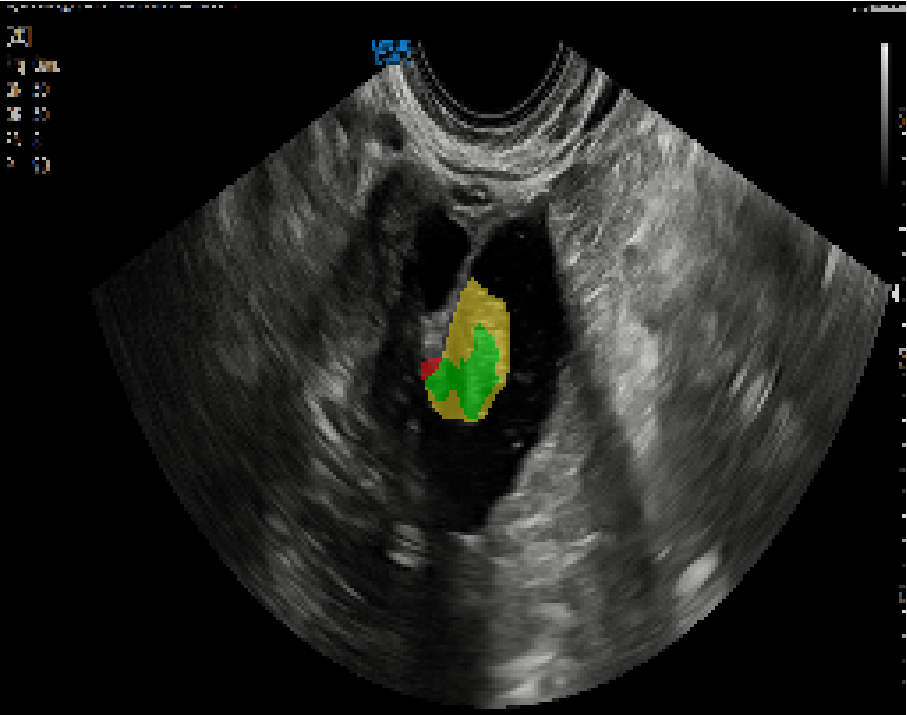}\\

\includegraphics[width=0.33\textwidth, height=40mm]{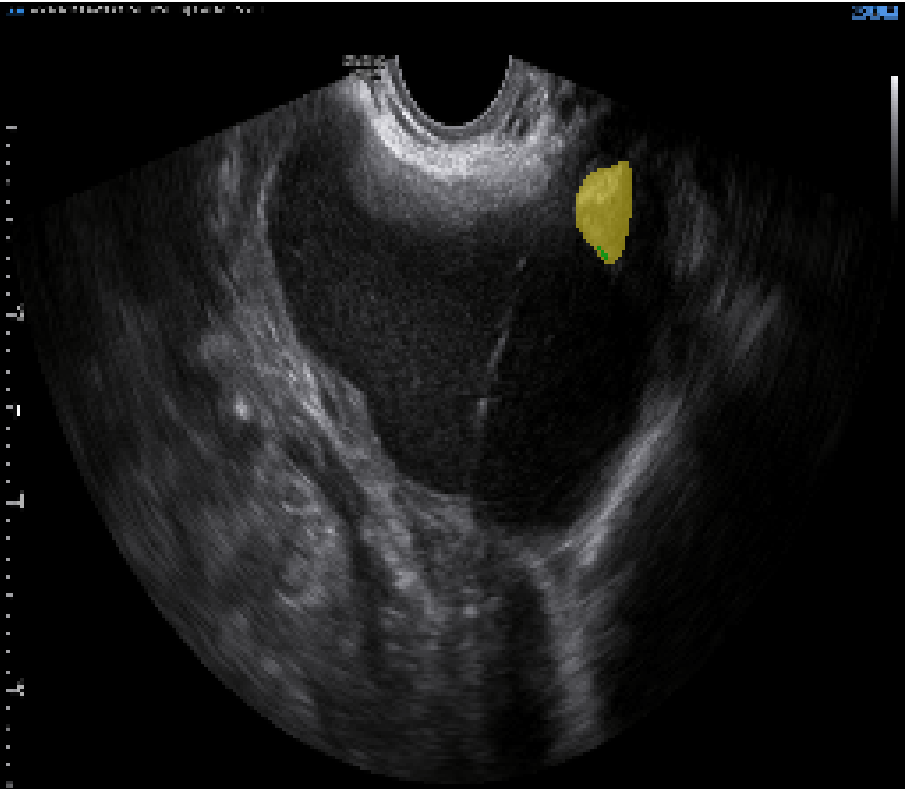} &
\includegraphics[width=0.33\textwidth, height=40mm]{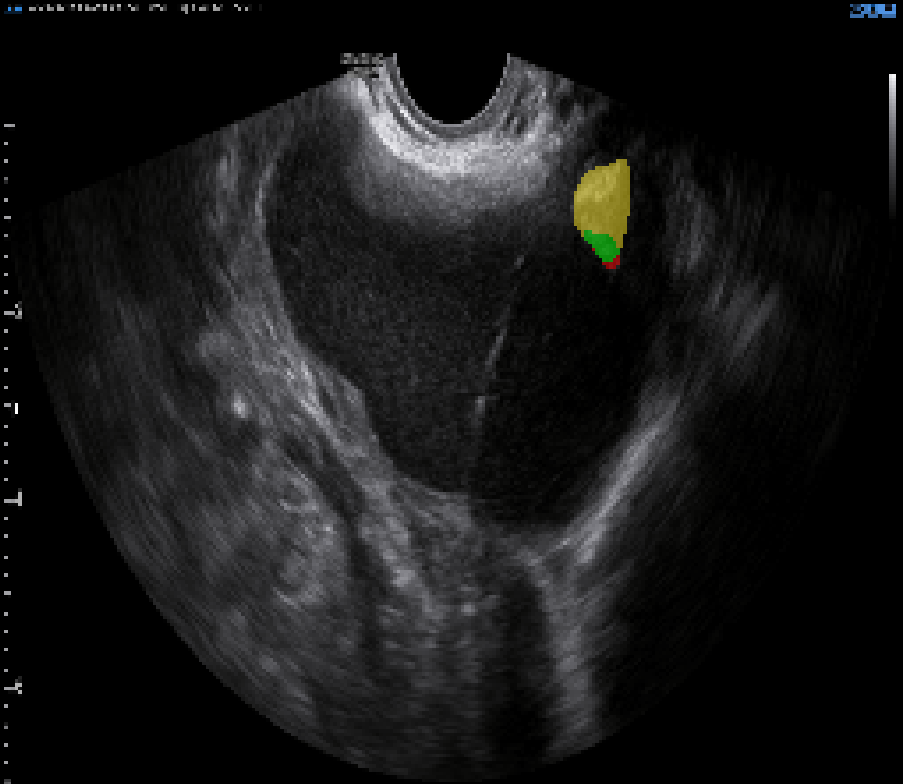} & 
\includegraphics[width=0.33\textwidth, height=40mm]{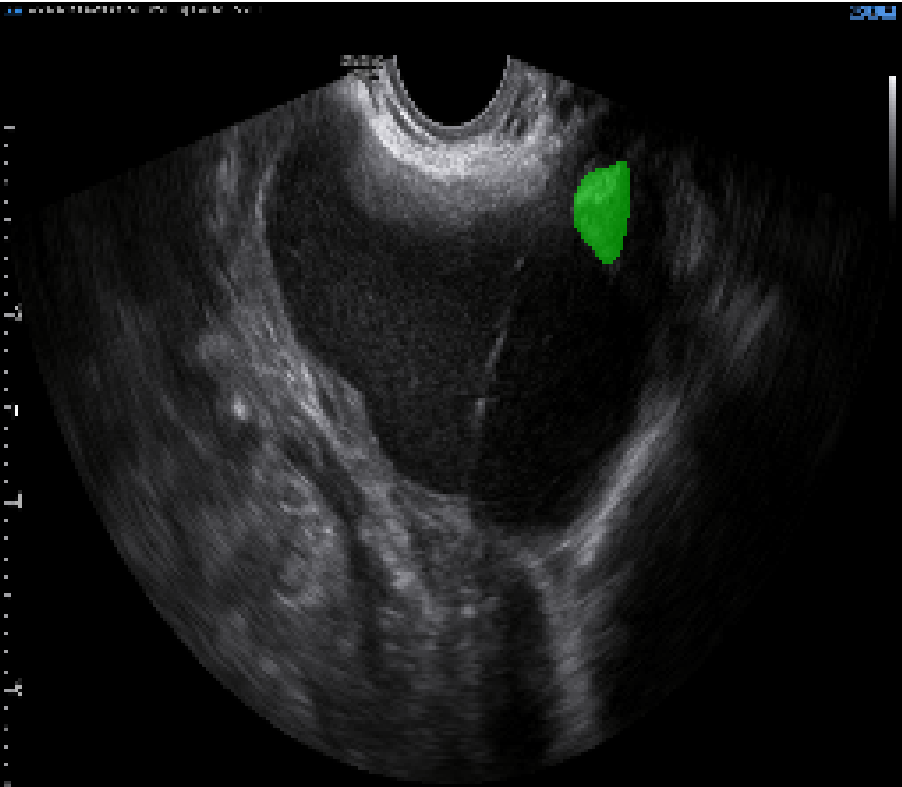} & 
\includegraphics[width=0.33\textwidth, height=40mm]{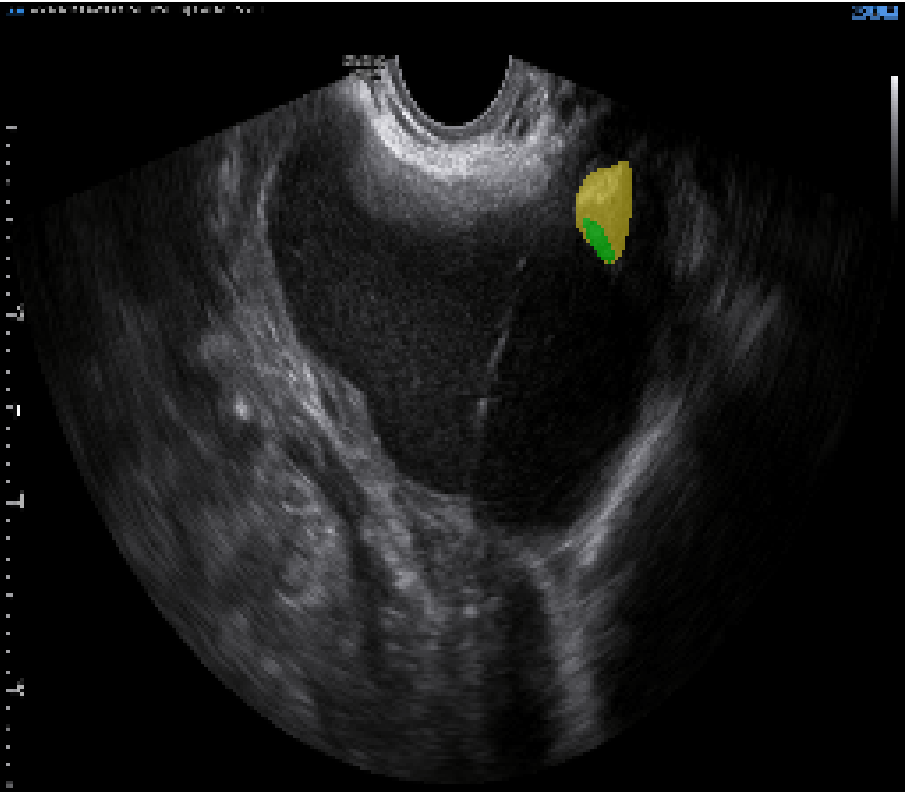} \\[.7cm] 

\multicolumn{4}{c}{\textbf{\Large Papillations}} \\[.4cm] 

\includegraphics[width=0.33\textwidth, height=40mm]{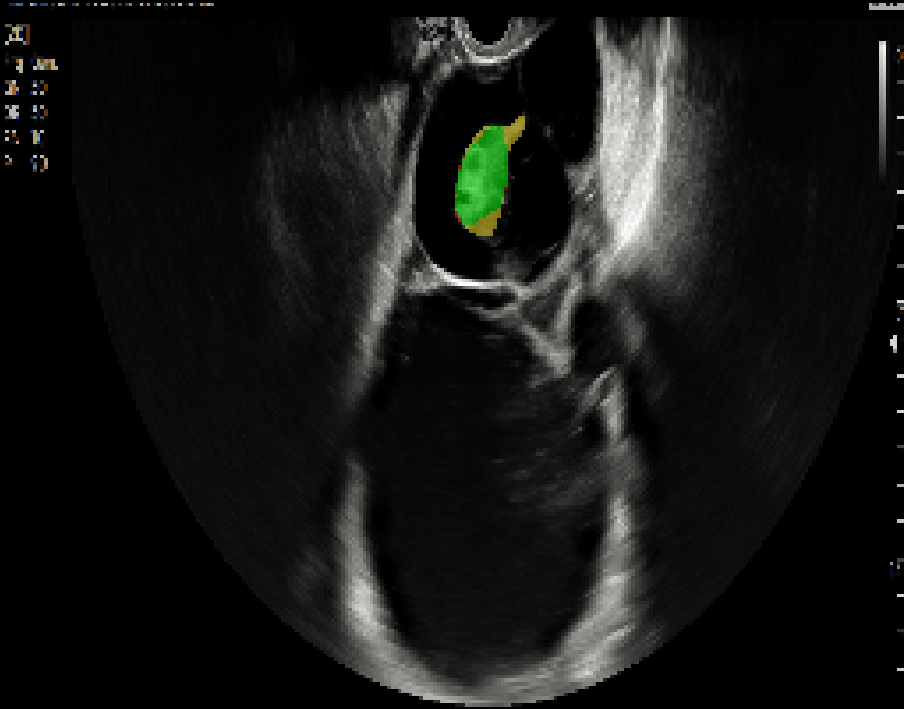} & 
\includegraphics[width=0.33\textwidth, height=40mm]{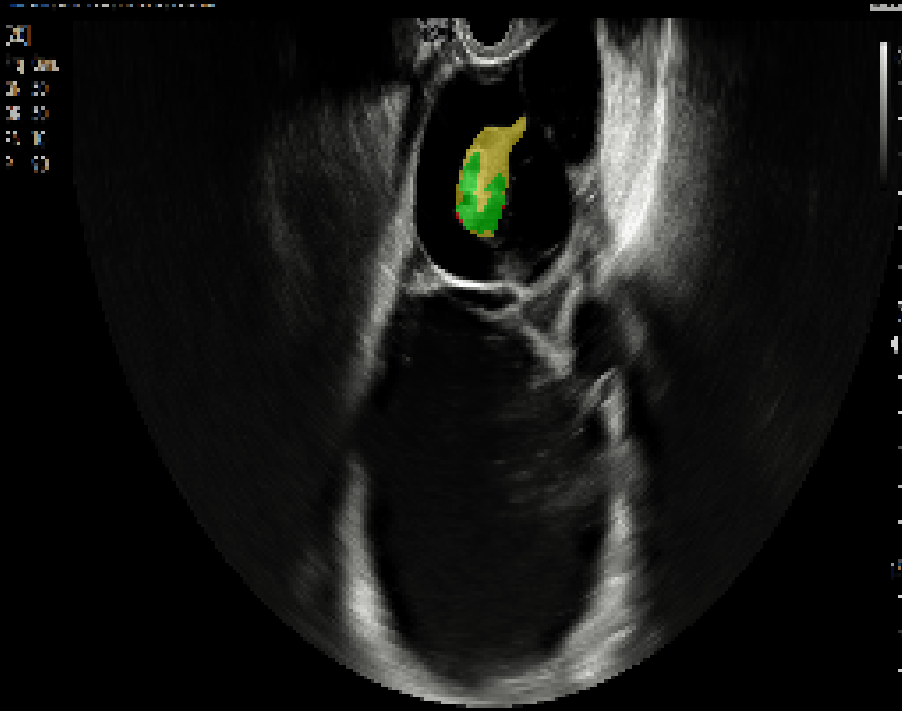} & 
\includegraphics[width=0.33\textwidth, height=40mm]{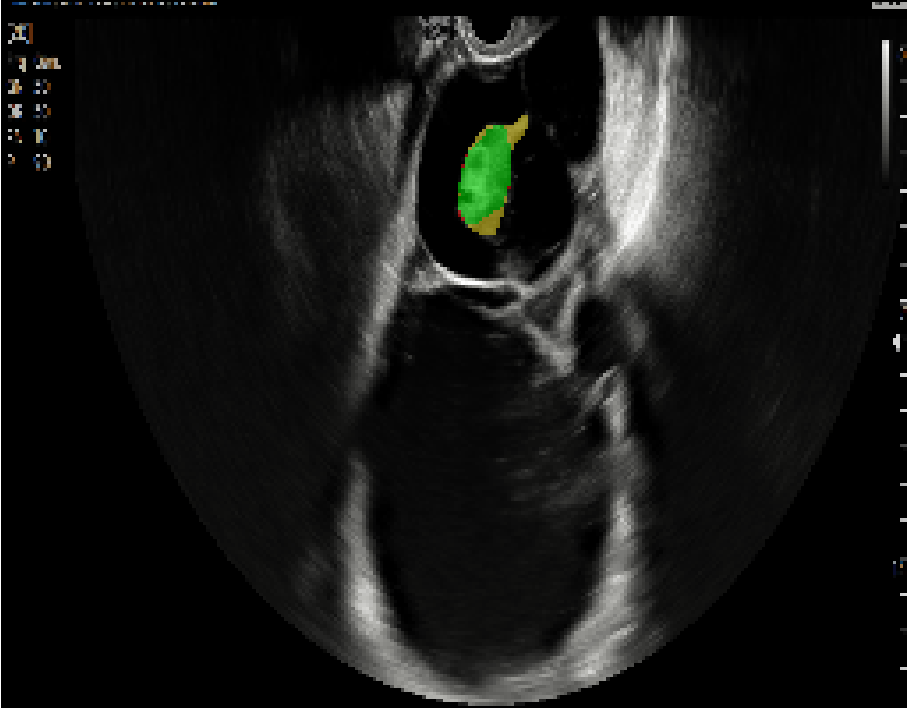} & 
\includegraphics[width=0.33\textwidth, height=40mm]{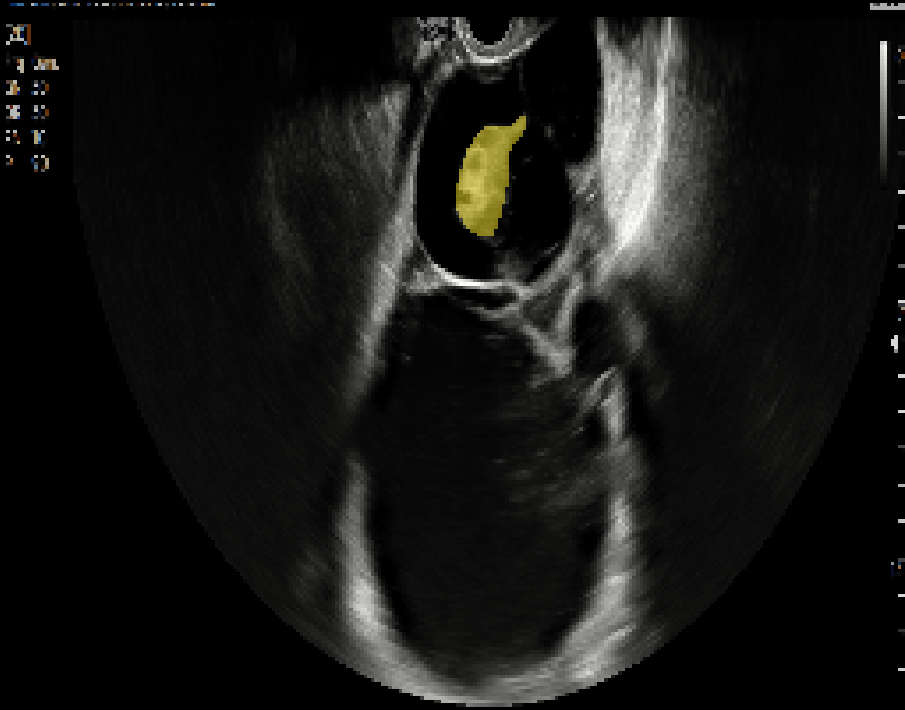}\\

\includegraphics[width=0.33\textwidth, height=40mm]{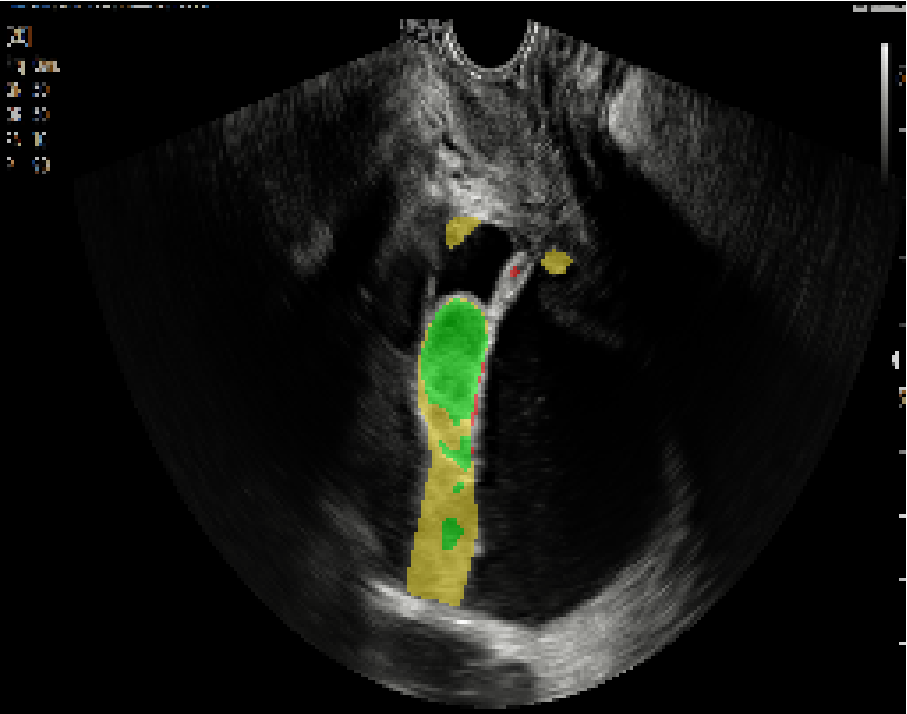} &
\includegraphics[width=0.33\textwidth, height=40mm]{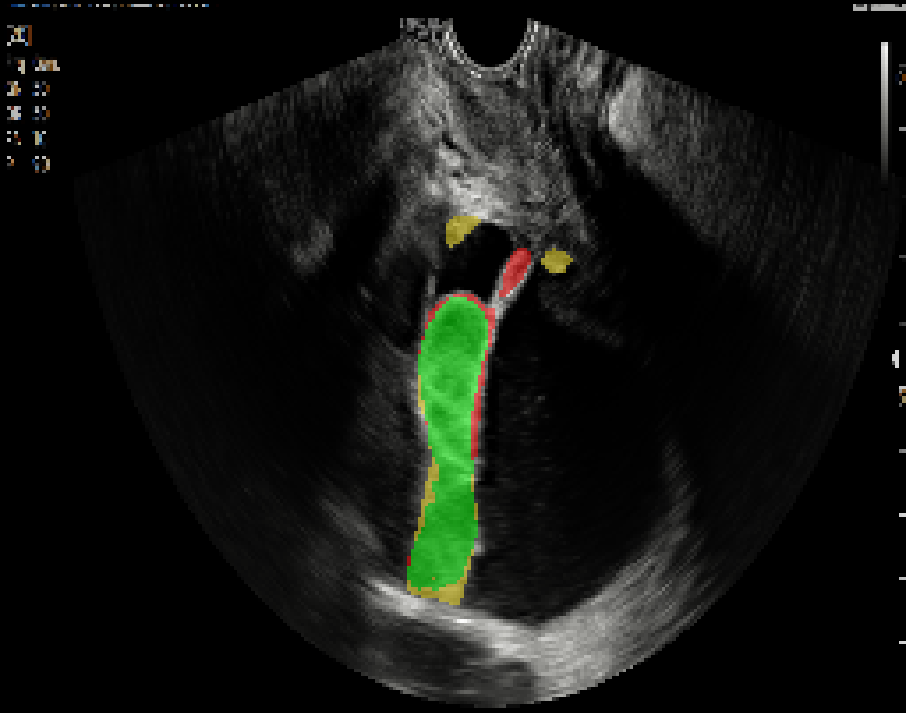} & 
\includegraphics[width=0.33\textwidth, height=40mm]{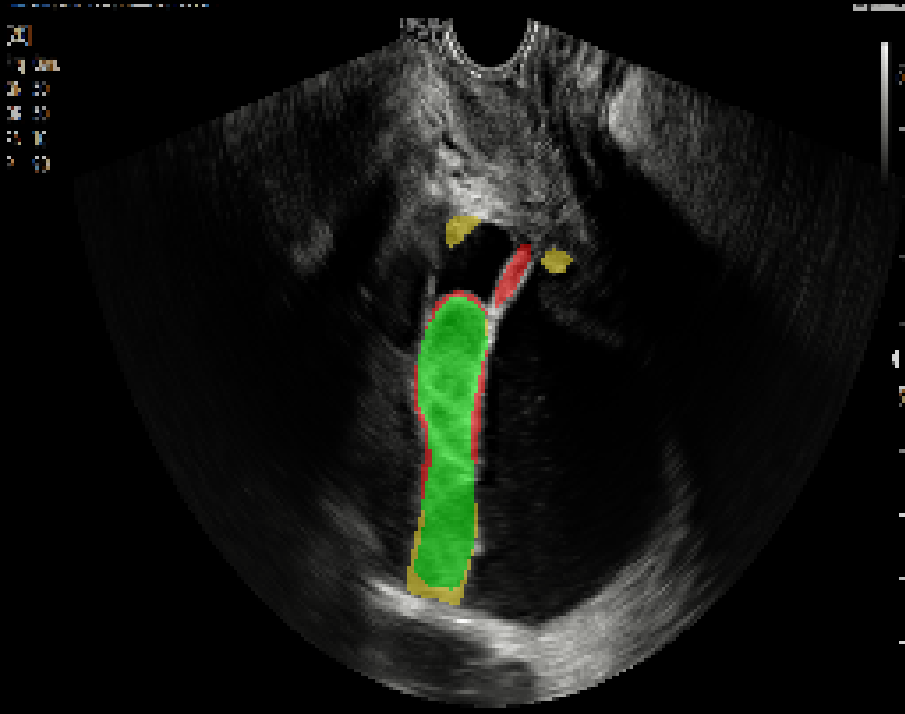} & \includegraphics[width=0.33\textwidth, height=40mm]{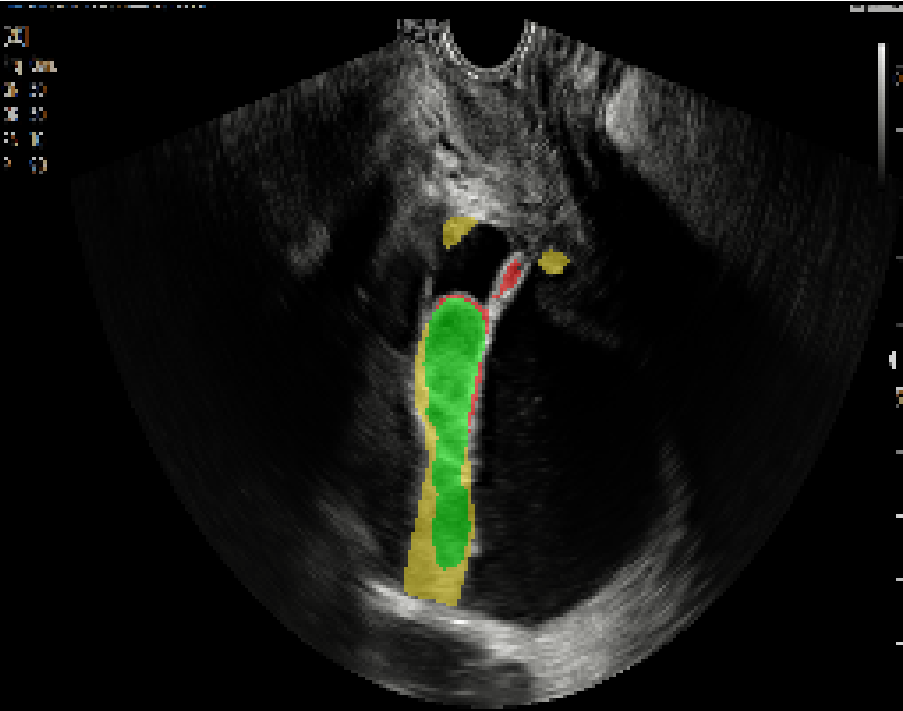} \\

\includegraphics[width=0.33\textwidth, height=40mm]{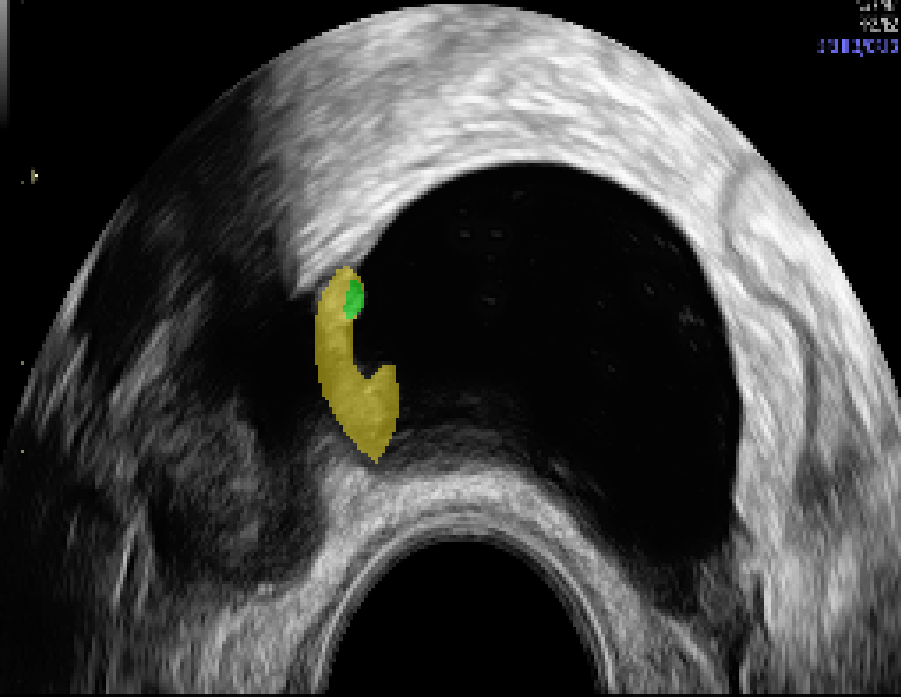} & \includegraphics[width=0.33\textwidth, height=40mm]{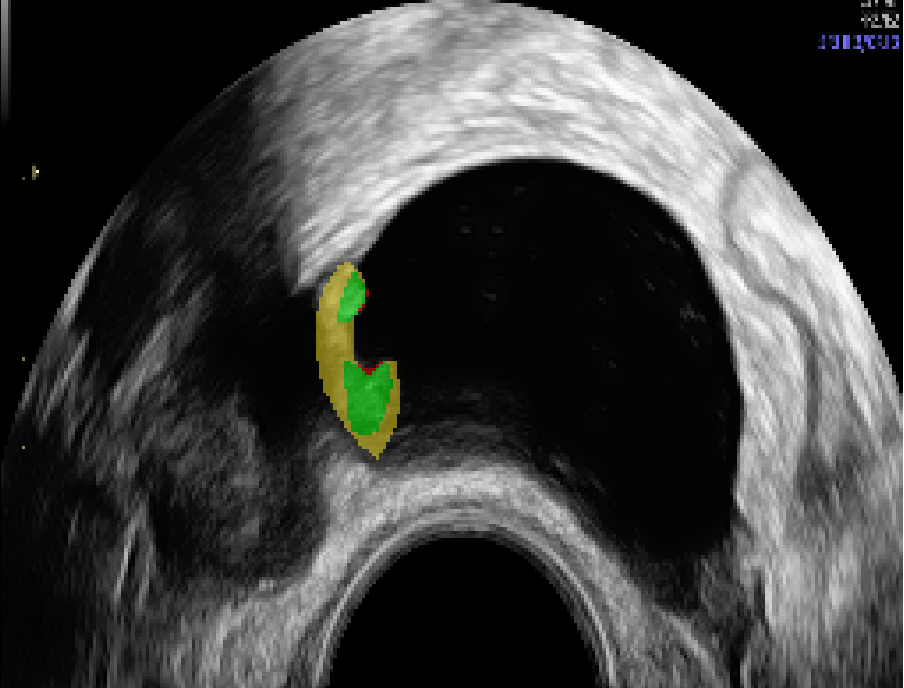} & 
\includegraphics[width=0.33\textwidth, height=40mm]{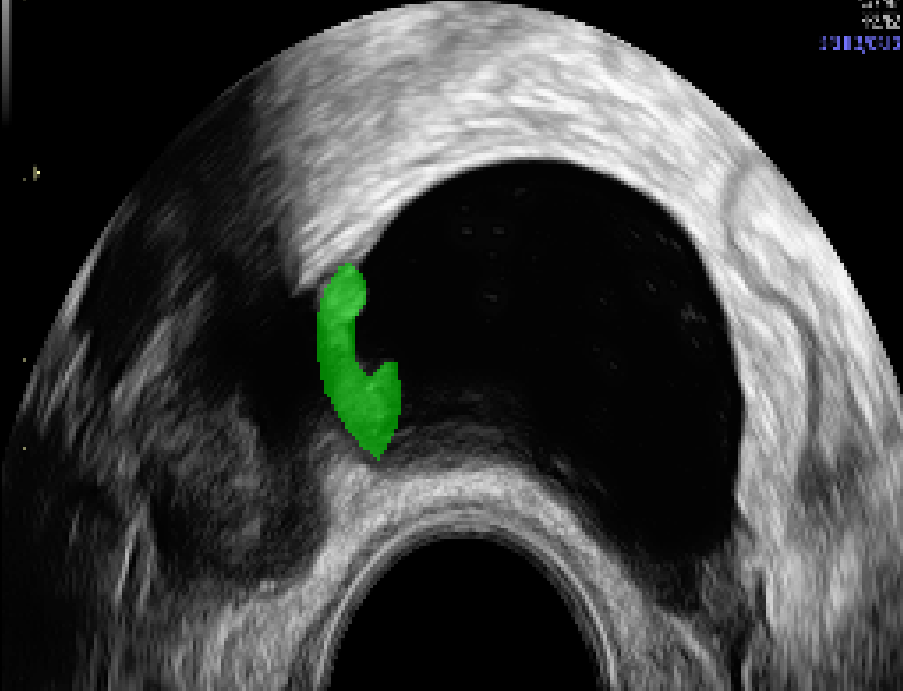} & 
\includegraphics[width=0.33\textwidth, height=40mm]{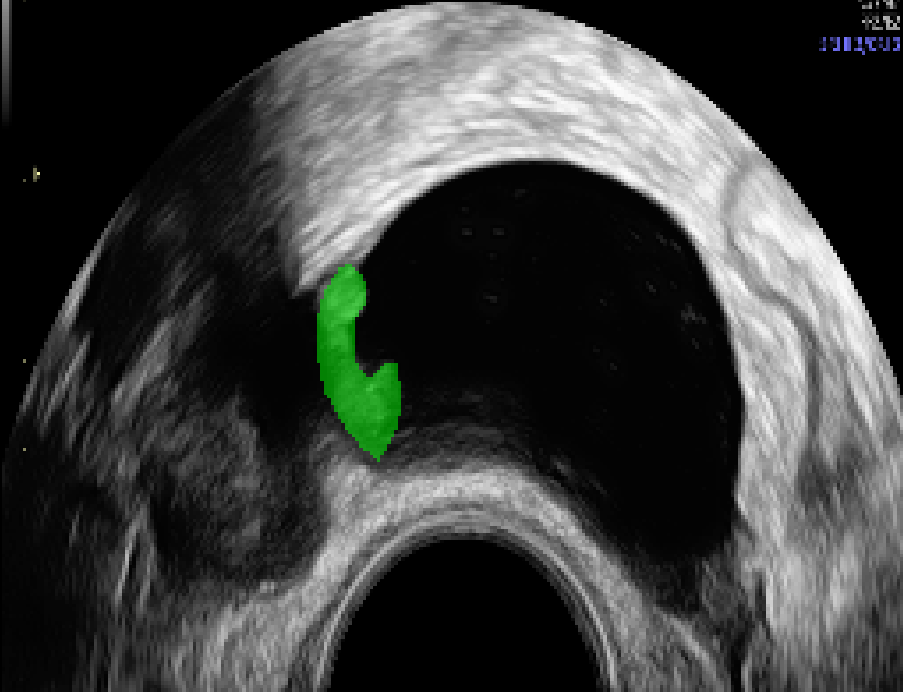} \\[.4cm]
\\
\textbf{\Large BL} & \textbf{\Large BL + BN + DA} & \textbf{\Large BL + S} & \textbf{\Large BL + S + BN + DA} \\
\\                        
\end{tabular}}
\caption{Predicted segmentations of trained models compared to the ground truth (GT) segmentation [\textbf{$\text{BL}$}: Baseline nnU-Net \cite{nn_unet} with region-based training, \textbf{BN}: Batch normalisation, \textbf{DA}: Intense data augmentation, \textbf{S}: With synthesised data] [Green represents correctly labelled pixels, red the falsely predicted segmentation and gold the false negative pixels]}
\label{tab:appendix:goodpredictionlabel3}
\end{figure}

\begin{figure}[htp]
\centering
\resizebox{1.0\textwidth}{!}{
  \begin{subfigure}[b]{0.25\textwidth}
    \includegraphics[height=3.0cm]{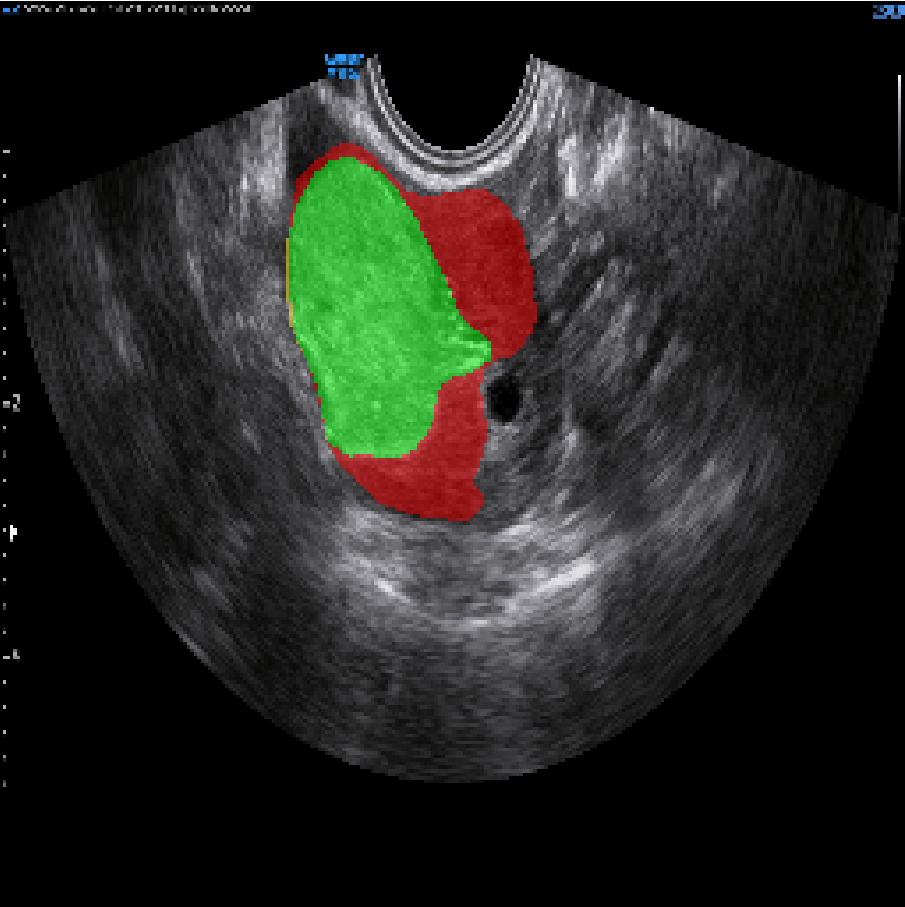}
    \caption{ %\newline \newline
    }
  \end{subfigure}
  \hfill
  \begin{subfigure}[b]{0.25\textwidth}
    \includegraphics[height=3.0cm]{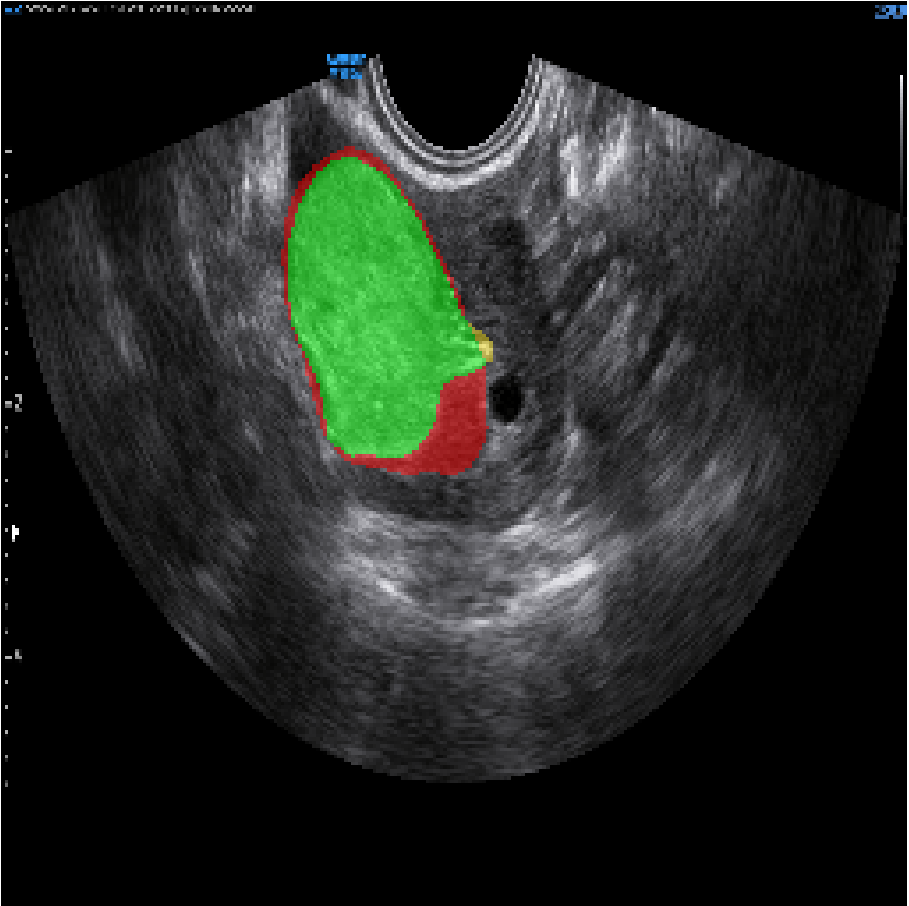}
    \caption{}
  \end{subfigure}
  \begin{subfigure}[b]{0.25\textwidth}
    \includegraphics[height=3.0cm]{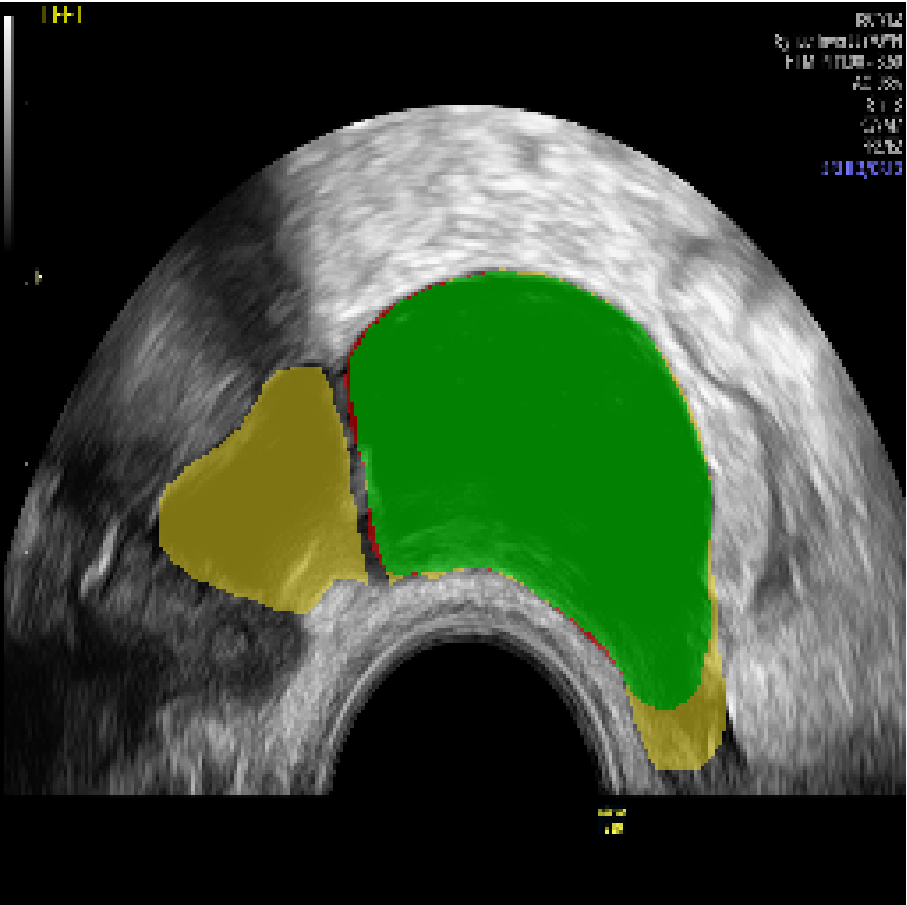}
    \caption{}
    \label{im:eval:multiloc1}
  \end{subfigure}
 % \hfill
  \begin{subfigure}[b]{0.25\textwidth}
    \includegraphics[height=3.0cm]{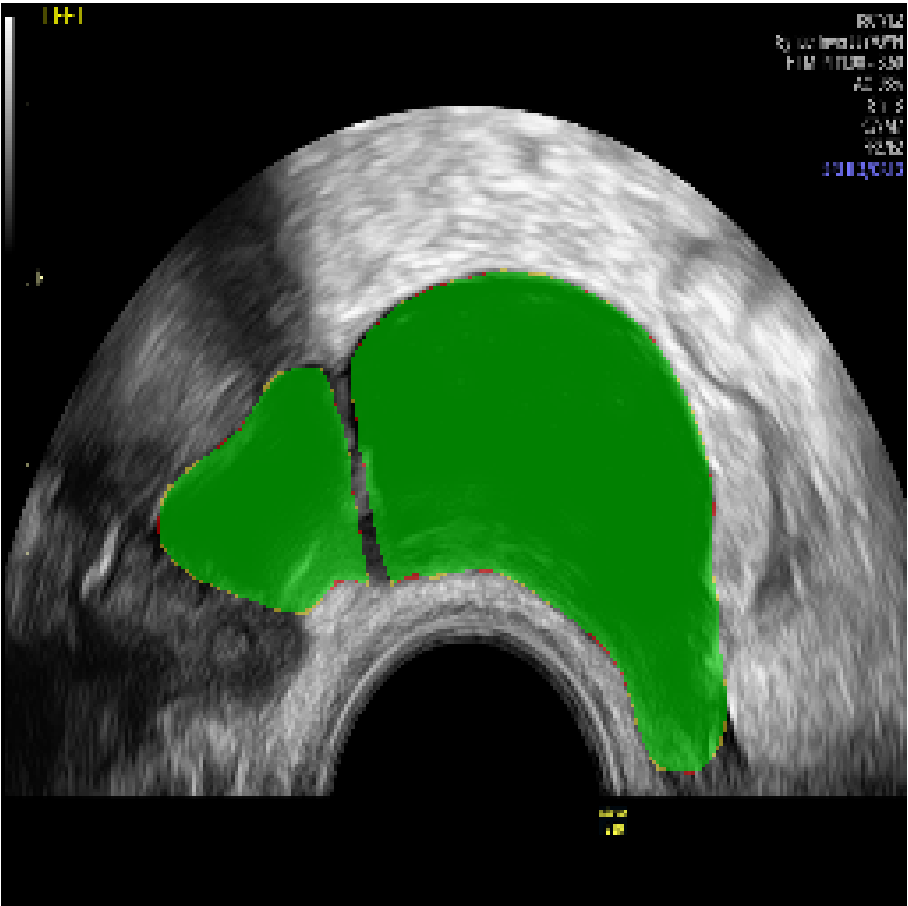}
    \caption{}
  \end{subfigure}}
  \caption{Sample of a segmentation of a lesion (a-b) and locule (c-d) [Green represents correctly labelled pixels, red the falsely predicted segmentation and gold the false negative pixels]. (a) Segmentation of a lesion by the baseline model. (b) Segmentation of a lesion by the baseline with synthesised data, batch normalisation and intense data augmentation model (BL + S + BN + DA). (c) Segmentation of locules by the baseline model, (d) Segmentation of locules by the baseline model with synthesised data.}
  \label{im:eval:seglabel1}
\end{figure}

The model with synthesised data was also trained with the addition of batch normalisation and intense data augmentation. The reason behind this was to apply more data augmentation onto the synthesised images and make the model more robust to all images and bridge the domain gap between the different machines used. The modifications appear to perform similarly to the baseline model with synthesised data. For solid areas, the performance is very similar to the model without BN and DA where the DSC and SDSC are still significantly better than the baseline with p < 0.01 and p < 0.05 respectively. For papillations, the DSC, SDSC and recall still remain within a single standard deviation, although the HD95 has increased. %The latter is most probably due to more papillations being segmented but causing a high number of false positives. 
Qualitative examples are shown in Figure~\ref{tab:appendix:goodpredictionlabel3} and \ref{im:eval:seglabel1}.  We used an Nvidia RTX A6000 for experimentation.

\noindent\textbf{Discussion and Limitations:}
%\textbf{Inter-observer variability}: 
Manual segmentation of ultrasound images can be error-prone, even when done by experts. The inter-observer variability of the DSC reaches a low of 0.771 with a variance of 0.4 for solid areas. Additionally, the disagreement rate is 20.8\% for papillations with a $\kappa$ of 0.17. Although the variance of our best performing model has decreased by 20\% (see Table \ref{tab:results}), we believe that having more than one step of verification for the ground truth segmentation would reduce the bias in our model further.
    
%\noindent\textbf{Synthesiser quality}:
We trained models with additional synthetic data on their own as well as combined with batch normalisation, data augmentation. The models with synthesised data perform best for all labels. In our analysis of the synthesised images, we established that most images looked realistic  (based on expert evaluation). However, a few images had blended papillations with clear edges and in some cases the direction of the speckle noise indicates artificial modifications. We hypothesise that improving the synthesiser and only including entirely realistic images would improve the performance further. There are also pathology constraints to be considered, \emph{e.g.}, a patient can only have a papillation if they have a solid area, a locule and a lesion, which represents only 10\% of our dataset. %This means that the dataset cannot be entirely balanced and therefore limits the potential of our synthesiser.
\iffalse
\section{Clinical Considerations}

This project is a medical research study which handles and analyses ultrasound images obtained from women. Our study recruited women who attended the Rapid Access and Ovarian Cyst clinics at the Hammersmith Hospital. The main requirement for our IOTA-7 study is for women to present with at least one adnexal mass. This study has been approved by the Queen Square Research Ethics Committee, and permission was granted for the use of ultrasound images for machine learning purposes. 

All patients have provided us with a written and signed consent form which, amongst several rights, allows us to use their scans as long as they are pseudo-anonymised. To do so, we have labelled their scans with a study number for training purposes, making the patients non-identifiable, and the scans are exported from the machines anonymously. Finally, no information directly identifying the patients has been used as part of the study.
\fi
\section{Conclusion}

This paper proposes a novel ultrasound image synthesiser to address dataset imbalance and under-represented adnexal masses. Our research has demonstrated that the nnU-Net baseline has room for improvements, specifically for masses with under-represented features: solid areas and papillations.
Medical experts have confirmed the realistic appearance of the synthesised images. The use of synthesised images balanced the dataset for papillations and solid areas, translating to an improvement in the performance of the model by 3.7\% and 8\% for papillations and solid areas respectively over the state-of-the-art nnU-Net ~\cite{nn_unet}. Furthermore, the integration of synthesised images also reduced the variability found between medical experts (by 35\% on average). 

\noindent\textbf{Acknowledgements:}  UKRI London Medical Imaging and Artificial Intelligence Centre for Value Based Healthcare (No.104691, \url{https://www.aicentre.co.uk/}). JB was supported by an Imperial Health Charity Pre-Doc Fellowship award 2021-2022. CL got support from the Imperial Dept. Computing Corporate Partnership Programme~\url{https://cpp.doc.ic.ac.uk/}.  

% \subsubsection{Acknowledgements} Please place your acknowledgements at
% the end of the paper, preceded by an unnumbered run-in heading (i.e.
% 3rd-level heading).
%\newpage
\bibliographystyle{splncs04}
\bibliography{bibliography}

\begin{thebibliography}{10}
\providecommand{\url}[1]{\texttt{#1}}
\providecommand{\urlprefix}{URL }
\providecommand{\doi}[1]{https://doi.org/#1}

\bibitem{budd2019confident}
Budd, S., Sinclair, M., Khanal, B., Matthew, J., Lloyd, D., Gomez, A.,
  Toussaint, N., Robinson, E.C., Kainz, B.: Confident head circumference
  measurement from ultrasound with real-time feedback for sonographers. In:
  International Conference on Medical Image Computing and Computer-Assisted
  Intervention. pp. 683--691. Springer (2019)

\bibitem{chotzoglou2021learning}
Chotzoglou, E., Day, T., Tan, J., Matthew, J., Lloyd, D., Razavi, R., Simpson,
  J., Kainz, B., et~al.: Learning normal appearance for fetal anomaly
  screening: Application to the unsupervised detection of hypoplastic left
  heart syndrome. Machine Learning for Biomedical Imaging  \textbf{1}(September
  2021 issue),  1--10 (2021)

\bibitem{chudecka2015roma}
Chudecka-G{\l}az, A.M.: Roma, an algorithm for ovarian cancer. Clinica Chimica
  Acta  \textbf{440},  143--151 (2015)

\bibitem{ovarian_cancer_stats_2}
Colombo, N., {Van Gorp}, T., Parma, G., Amant, F., Gatta, G., Sessa, C.,
  Vergote, I.: Ovarian cancer. Critical Reviews in Oncology/Hematology
  \textbf{60}(2),  159--179 (2006).
  \doi{https://doi.org/10.1016/j.critrevonc.2006.03.004}

\bibitem{cronin2020using}
Cronin, N.J., Finni, T., Seynnes, O.: Using deep learning to generate synthetic
  b-mode musculoskeletal ultrasound images. Computer methods and programs in
  biomedicine  \textbf{196},  105583 (2020)

\bibitem{gholami2018novel}
Gholami, A., Subramanian, S., Shenoy, V., Himthani, N., Yue, X., Zhao, S., Jin,
  P., Biros, G., Keutzer, K.: A novel domain adaptation framework for medical
  image segmentation. In: International MICCAI Brainlesion Workshop. pp.
  289--298. Springer (2018)

\bibitem{hesamian2019deep}
Hesamian, M.H., Jia, W., He, X., Kennedy, P.: Deep learning techniques for
  medical image segmentation: achievements and challenges. Journal of digital
  imaging  \textbf{32}(4),  582--596 (2019)

\bibitem{nn_unet}
Isensee, F., Jaeger, P.F., Kohl, S.A., Petersen, J., Maier-Hein, K.H.: nnu-net:
  a self-configuring method for deep learning-based biomedical image
  segmentation. Nature methods  \textbf{18}(2),  203--211 (2021)

\bibitem{jiao2020self}
Jiao, J., Namburete, A.I., Papageorghiou, A.T., Noble, J.A.: Self-supervised
  ultrasound to mri fetal brain image synthesis. IEEE Transactions on Medical
  Imaging  \textbf{39}(12),  4413--4424 (2020)

\bibitem{jin2021multiple}
Jin, J., Zhu, H., Zhang, J., Ai, Y., Zhang, J., Teng, Y., Xie, C., Jin, X.:
  Multiple u-net-based automatic segmentations and radiomics feature stability
  on ultrasound images for patients with ovarian cancer. Frontiers in Oncology
  \textbf{10}, ~3428 (2021)

\bibitem{746626}
Krivanek, A., Sonka, M.: Ovarian ultrasound image analysis: follicle
  segmentation. IEEE Transactions on Medical Imaging  \textbf{17}(6),  935--944
  (1998). \doi{10.1109/42.746626}

\bibitem{kumar2014segmentation}
Kumar, H.P., Srinivasan, S.: Segmentation of polycystic ovary in ultrasound
  images. In: Second International Conference on Current Trends In Engineering
  and Technology-ICCTET 2014. pp. 237--240. IEEE (2014)

\bibitem{meng2020mutual}
Meng, Q., Matthew, J., Zimmer, V.A., Gomez, A., Lloyd, D.F., Rueckert, D.,
  Kainz, B.: Mutual information-based disentangled neural networks for
  classifying unseen categories in different domains: Application to fetal
  ultrasound imaging. IEEE transactions on medical imaging  \textbf{40}(2),
  722--734 (2020)

\bibitem{meng2019weakly}
Meng, Q., Sinclair, M., Zimmer, V., Hou, B., Rajchl, M., Toussaint, N., Oktay,
  O., Schlemper, J., Gomez, A., Housden, J., et~al.: Weakly supervised
  estimation of shadow confidence maps in fetal ultrasound imaging. IEEE
  transactions on medical imaging  \textbf{38}(12),  2755--2767 (2019)

\bibitem{nhs_cancer_stat}
NHS: Overview - cancer (2022), \url{https://www.nhs.uk/conditions/cancer/}

\bibitem{opencvpoisson}
OpenCV: Opencv documentation (2022),
  \url{https://docs.opencv.org/4.x/df/da0/group__photo__clone.html}

\bibitem{perez2003poisson}
P\'{e}rez, P., Gangnet, M., Blake, A.: Poisson image editing. In: ACM SIGGRAPH
  2003 Papers. p. 313–318. SIGGRAPH '03, Association for Computing Machinery,
  New York, NY, USA (2003). \doi{10.1145/1201775.882269},
  \url{https://doi.org/10.1145/1201775.882269}

\bibitem{roscoecoping}
Roscoe, J.F., Dee, H., Zwiggelaar, R.: Coping with noise in ultrasound images:
  A review. Medical Image Understanding and Analysis: 25th Annual Conference,
  MIUA 2021

\bibitem{no2011management}
{Royal College of Obstetricians Gynaecologists}, Green-top Guideline, N..:
  Management of suspected ovarian masses in premenopausal women  (2016)

\bibitem{shin2018medical}
Shin, H.C., Tenenholtz, N.A., Rogers, J.K., Schwarz, C.G., Senjem, M.L.,
  Gunter, J.L., Andriole, K.P., Michalski, M.: Medical image synthesis for data
  augmentation and anonymization using generative adversarial networks. In:
  International workshop on simulation and synthesis in medical imaging. pp.
  1--11. Springer (2018)

\bibitem{tan2022detecting}
Tan, J., Hou, B., Batten, J., Qiu, H., Kainz, B., et~al.: Detecting outliers
  with foreign patch interpolation. Machine Learning for Biomedical Imaging
  \textbf{1}(April 2022 issue),  1--10 (2022)

\bibitem{timmerman2005logistic}
Timmerman, D., Testa, A.C., Bourne, T., Ferrazzi, E., Ameye, L.,
  Konstantinovic, M.L., Van~Calster, B., Collins, W.P., Vergote, I.,
  Van~Huffel, S., et~al.: Logistic regression model to distinguish between the
  benign and malignant adnexal mass before surgery: a multicenter study by the
  international ovarian tumor analysis group. Journal of Clinical Oncology
  \textbf{23}(34),  8794--8801 (2005)

\bibitem{timmerman2008simple}
Timmerman, D., Testa, A.C., Bourne, T., Ameye, L., Jurkovic, D., Van~Holsbeke,
  C., Paladini, D., Van~Calster, B., Vergote, I., Van~Huffel, S., et~al.:
  Simple ultrasound-based rules for the diagnosis of ovarian cancer. Ultrasound
  in Obstetrics and Gynecology: The Official Journal of the International
  Society of Ultrasound in Obstetrics and Gynecology  \textbf{31}(6),  681--690
  (2008)

\bibitem{cancerresearch}
UK, C.R.: Ovarian cancer statistics (2022),
  \url{https://www.cancerresearchuk.org/health-professional/cancer-statistics/statistics-by-cancer-type/ovarian-cancer#heading-One}

\bibitem{van2014evaluating}
Van~Calster, B., Van~Hoorde, K., Valentin, L., Testa, A.C., Fischerova, D.,
  Van~Holsbeke, C., Savelli, L., Franchi, D., Epstein, E., Kaijser, J., et~al.:
  Evaluating the risk of ovarian cancer before surgery using the adnex model to
  differentiate between benign, borderline, early and advanced stage invasive,
  and secondary metastatic tumours: prospective multicentre diagnostic study.
  Bmj  \textbf{349} (2014)

\bibitem{RMI_comparison}
Yamamoto, Y., Tsuchida, A., Ushiwaka, T., Nagai, R., Matsumoto, M., Komatsu,
  J., Kinoshita, H., Minami, S., Hayashi, K.: Comparison of 4
  risk-of-malignancy indexes in the preoperative evaluation of patients with
  pelvic masses: A prospective study. Clinical Ovarian and Other Gynecologic
  Cancer  \textbf{7} (11 2014). \doi{10.1016/j.cogc.2014.11.001}

\end{thebibliography}
\newpage

\end{document}